\documentclass[11pt]{article}
 
\def\fl{\noindent}
\newcommand{\bra}{\langle}
\newcommand{\ket}{\rangle}
\newcommand{\lv}{\left\vert}
\newcommand{\rv}{\right\vert}
\newcommand{\tl}[1]{\tilde{#1}}
\newcommand{\pdr}{\partial}
\newcommand{\grad}{{\bf \nabla}}
\newcommand{\beq}{\begin{equation}}
\newcommand{\eeq}{\end{equation}}
\newcommand{\beqs}{\begin{eqnarray}}
\newcommand{\eeqs}{\end{eqnarray}}
\newcommand{\half}{\frac{1}{2}}
\newcommand{\ov}[1]{\frac{1}{#1}}
\newcommand{\fr}[2]{\frac{#1}{#2}}

		\def\g{\gamma} 		\def\G{\Gamma} 		\def\del{\delta}	\def\D{\Delta}		 	\def\la{\lambda}			\def\tht{\theta}	\def\om{\omega}		

\usepackage{amsmath,amssymb}

\usepackage[pdftex]{hyperref,color,graphicx}

\usepackage{calligra} 
\DeclareMathAlphabet{\mathcalligra}{T1}{calligra}{m}{n}
\DeclareFontShape{T1}{calligra}{m}{n}{<->s*[2.2]callig15}{}
\newcommand{\scripty}[1]{\ensuremath{\mathcalligra{#1}}}

\newcount\colveccount  
\newcommand*\colvec[1]{\global\colveccount#1  \begin{pmatrix} \colvecnext} \def\colvecnext#1{#1 \global\advance\colveccount-1
        \ifnum\colveccount>0 \\ \expandafter\colvecnext
        \else \end{pmatrix} \fi}

\newcommand{\bfr}{{\bf r}}

\newcommand{\bfp}{{\bf p}}

\newcommand{\mC}{{\mathbb{C}}}
\newcommand{\mR}{{\mathbb{R}}}
\newcommand{\mS}{{\mathbb{S}}}

\usepackage{subcaption}

\begin{document}



\title{\normalsize 
\hfill {\tt arXiv:1606.05091} \\
\vskip 0.1mm \LARGE
Curvature and geodesic instabilities in a geometrical approach to the planar three-body problem
\\
}
\author{{\sc Govind S. Krishnaswami and Himalaya Senapati}
\\ \\ \small
 Physics Department, Chennai Mathematical Institute,  SIPCOT IT Park, Siruseri 603103, India
\\ \small
 Email: {\tt govind@cmi.ac.in, himalay@cmi.ac.in}}

\date{Oct 8, 2016 \\ Published in J. Math. Phys. 57, 102901 (2016)}

\maketitle

\begin{abstract} \normalsize

The Maupertuis principle allows us to regard classical trajectories as reparametrized geodesics of the Jacobi-Maupertuis (JM) metric on configuration space. We study this geodesic reformulation of the {\it planar} three-body problem with both Newtonian and attractive inverse-square potentials. The associated JM metrics possess translation and rotation isometries in addition to scaling isometries for the inverse-square potential with zero energy $E$. The geodesic flow on the {\it full} configuration space $\mathbb{C}^3$ (with collision points excluded) leads to corresponding flows on its Riemannian quotients: the center of mass configuration space $\mathbb{C}^2$ and shape space $\mathbb{R}^3$ (as well as $\mathbb{S}^3$ and the shape sphere $\mathbb{S}^2$ for the inverse-square potential when $E = 0$). The corresponding Riemannian submersions are described explicitly in `Hopf' coordinates which are particularly adapted to the isometries. For equal masses subject to inverse-square potentials, Montgomery shows that the zero-energy `pair of pants' JM metric on the shape sphere is geodesically complete and has negative gaussian curvature except at Lagrange points. We extend this to a proof of boundedness and strict negativity of scalar curvatures everywhere on $\mathbb{C}^2$, $\mathbb{R}^3$ and $\mathbb{S}^3$ with collision points removed. Sectional curvatures are also found to be largely negative, indicating widespread geodesic instabilities. We obtain asymptotic metrics near collisions, show that scalar curvatures have finite limits and observe that the geodesic reformulation `regularizes' pairwise and triple collisions on $\mathbb{C}^2$ and its quotients for arbitrary masses and allowed energies. For the Newtonian potential with equal masses and zero energy, we find that the scalar curvature on $\mathbb{C}^2$ is strictly negative though it could have either sign on $\mathbb{R}^3$. However, unlike for the inverse-square potential, geodesics can encounter curvature singularities at collisions in finite geodesic time.


\end{abstract} \normalsize




 



\scriptsize

\tableofcontents

\normalsize

\section{Introduction}

The classical three-body problem and associated questions of stability have stimulated much work in mechanics and nonlinear \& chaotic dynamics \cite{gutzwiller-three-body,chenciner-poincare-three-body,routh,laskar,montgomery-notices-ams}. Quantum and fluid mechanical variants with potentials other than Newtonian are also of interest: e.g., the dynamics of two-electron atoms and the water molecule \cite{gutzwiller-book}, the $N$-vortex problem with logarithmic potentials \cite{newton-n-vortex}, the problem of three identical bosons with inverse-square potenials (Efimov effect in cold atoms \cite{efimov, efimov-cold-atom}) and the Calogero-Moser system also with inverse-square potentials \cite{calogero}. We investigate a geometrical approach to the planar three-body problem with Newtonian and attractive inverse-square potentials. The inverse-square potential has some simplifying features over the Newtonian one due in part to the nature of its scaling symmetry $H(\la \bfr, \la^{-1} \bfp) = \la^{-2} H(\bfr, \bfp)$. As a consequence, the sign of energy $E$ controls asymptotic behaviour: bodies fly apart or suffer a triple collision according as $E$ is positive/negative, leaving open the special case $E=0$ \cite{Rajeev}. This follows from the Lagrange-Jacobi identity $\ddot I = 4 E$ for the evolution of the moment of inertia $I = \sum m_i {\bf r}_i^2$. By contrast, for the Newtonian potential, $H(\la^{-2/3} \bfr, \la^{1/3} \bfp) = \la^{2/3} H(\bfr, \bfp)$ leads to $\ddot I = 4E - 2V$, which is not sufficient to determine the long time behavior of $I$ when $E < 0$.

Our approach is based on a geometric reformulation of Newtonian trajectories.
It is well known that trajectories of a free particle moving on a Riemannian manifold are geodesics of a mass/kinetic metric $m_{ij}$ defined by the kinetic energy $\half m_{ij}(x) \dot x^i \dot x^j$. Indeed, geodesic flow on a compact Riemann surface of constant negative curvature is a prototypical model for chaos \cite{gutzwiller-book}. In the presence of a potential $V$, trajectories are reparametrized geodesics of the conformally related Jacobi-Maupertuis (JM) metric $g_{ij} = (E-V(x))m_{ij}$ (see Refs.\cite{lanczos, arnold} and \S \ref{s:traj-as-geodesics}). The linear stability of geodesics to perturbations is then controlled by sectional curvatures of the JM metric.

Several authors have tried to relate the geometry of the JM metric to chaos. For systems with many degrees of freedom, Pettini et. al. \cite{pettini-2000-phys-rpts,pettini-book-2007,pettini-2008} obtain an approximate expression for the largest Lyapunov exponent in terms of curvatures. In Ref. \cite{pettini-1996} the geometric framework is applied to investigate chaos in the H\'enon-Heiles system and a suitable average sectional curvature proposed as an indicator of chaos for systems with few degrees of freedom (see also \cite{Ramasubramanian-Sriram}). While negativity of curvature need not imply chaos, as the Kepler problem shows for $E > 0$, these works suggest that chaos could arise both from negativity of curvature and from fluctuations in curvature through parametric instabilities.

For the {\it planar} gravitational three-body problem (i.e. with pairwise Newtonian potentials), the JM metric on the full configuration space $\mR^6 \cong \mC^3$ has isometries corresponding to translation and rotation invariance groups $\bf C$ and U$(1)$ (\S \ref{s:jm-metric-config-space-hopf-coords}). This allows one to study the reduced dynamics on the quotients: configuration space $\mC^2 \cong \mC^3/{\bf C}$ and shape space $\mR^3 \cong \mC^2/{\rm U}(1)$ \cite{montgomery-american-monthly}. Here, collision configurations are excluded from $\mC^3$ and its quotients. When the Newtonian potential is replaced with the inverse-square potential, the zero-energy JM metric  acquires a scaling isometry leading to additional quotients: $\mS^3 \cong \mC^2/{\rm scaling}$ and the shape sphere $\mS^2 \cong \mR^3/{\rm scaling}$ (see Fig. \ref{f:flow-chart}). Since the three collision points have been removed, the (non-compact) shape sphere $\mS^2$ has the topology of a pair of pants and fundamental group given by the free group on two generators. As part of a series of works on the planar three-body problem, Montgomery \cite{montgomery-pants} shows that for three equal masses with inverse-square potentials (sometimes referred to as a `strong' force), the curvature of the JM metric on $\mS^2$ is negative except at the two Lagrange points, where it vanishes. As a corollary, he shows the uniqueness of the `figure $8$' solution and establishes that collision solutions are dense within bound ones. In Ref. \cite{montgomery-syzygy,montgomery-2007}, he uses the geometry of the shape sphere to show that  zero angular momentum negative energy solutions (other than the Lagrange homotheties) of the gravitational three-body problem have at least one syzygy (collinearity).

In this paper, we begin by extending some of Montgomery's results on the geometry of the shape sphere to that of the configuration space $\mC^2$ (without any restriction on angular momentum) and its quotients. Metrics on the quotients are obtained explicitly via Riemannian submersions (\S \ref{s:quotient-metrics}, \S \ref{s:curvature-newtonian-potential}) which simplify in `Hopf' coordinates, as the Killing vector fields (KVFs) point along coordinate vector fields. These coordinates also facilitate our explicit computation of metrics and curvatures near binary and triple collisions. We interpret Lagrange and Euler homotheties (`central configurations' \cite{chenciner-scholarpedia}) as radial geodesics at global and local minima of the conformal factor in the JM metric for the inverse-square potential (\S \ref{s:geodesic-completeness}) and thereby deduce geodesic completeness of the configuration manifold $\mC^2$ and its quotients $\mR^3$ and $\mS^3$ for arbitrary masses and allowed energies. The estimates showing completeness on $\mC^2$ are similar to those showing that the classical action (integral of Lagrangian) diverges for collisional trajectories. In a private communication, R Montgomery points out that this was known to Poincare and has been rediscovered several times (see for example Ref. \cite{c-moore-braid-group, montgomery-braid-groups, chenciner-icm-notes}). Completeness establishes that the geodesic reformulation `regularizes' pairwise and triple collisions by reparametrizing time so that any collision occurs at $t = \infty$. In contrast with other regularizations \cite{yeomans,celletti}, this does not involve an extrapolation of the dynamics past a collision nor a change in dependent variables. Unlike for the inverse-square potential, we show that geodesics for the Newtonian potential can reach curvature singularities (binary/triple collisions) in finite geodesic time (\S \ref{s:geodesic-incompleteness-newtonian-pot}). This may come as a surprise, since the Newtonian potential is {\it less} singular than the inverse-square potential and masses collide sooner under Newtonian evolution in the inverse-square potential. However, due to the reparametrization of time in going from trajectories to geodesics, masses can collide in finite time in the Newtonian potential while taking infinitely long to do so in the inverse-square potential. Indeed, for the attractive $1/r^n$ potential, the JM line-element leads to estimates $\propto \int_0^{\eta_0} \frac{d \eta}{\eta^{n/2}}$ and $\int_0^{r_0} \frac{d r}{r^{n/2}}$ for the distances to binary and triple collisions from a nearby location (\S \ref{s:geodesic-completeness}). These diverge for $n \geq 2$ and are finite for $n < 2$.

To examine stability of geodesics, we evaluate scalar and sectional curvatures of the zero-energy, equal-mass JM metrics on $\mC^2$ and its quotients. For the inverse-square potential, we obtain strictly negative upper bounds for scalar curvatures on $\mC^2$, $\mR^3$ and $\mS^3$ (\S \ref{s:scalar-curvature-inv-sq-pot-c2-r3-s3-s2}), indicating widespread linear geodesic instability. Moreover, scalar curvatures are shown to be bounded below. In particular, they remain finite and negative at binary and triple collisions. O'Neill's theorem is used to determine or bound various sectional curvatures on $\mC^2$ using the more easily determined ones on its Riemannian quotients; they are found to be largely negative (\S \ref{s:sectional-curvature-inv-sq-pot}). On the other hand, for the Newtonian potential, we find that the scalar curvature on $\mC^2$ is strictly negative, though it can have either sign on shape space $\mR^3$ (\S \ref{s:curvature-newtonian-potential}). Unlike for the inverse-square potential, scalar curvatures $\to - \infty$ at collision points. We also discuss the geodesic instability of Lagrange rotation and homothety solutions for equal masses (\S \ref{s:stability-tensor}). We end with a cautionary remark  comparing stability of geodesics to that of corresponding trajectories, simple examples are used to illustrate that the two notions of stability need not always coincide. In this paper we have not touched upon the interesting issues of long-term geodesic stability or chaos. It would be interesting to relate the local geodesic instabilities discussed here to medium- and long-time behavior. The dynamical consequences of sectional curvatures possessing either sign should also be of much interest.

\section{Trajectories as geodesics of the Jacobi-Maupertuis metric}
\label{s:traj-as-geodesics}

For a system with configuration space $M$ and Lagrangian $L = (1/2) m_{ij}(x) \dot x^i \dot x^j$, Lagrange's equations are equivalent to the geodesic equations with respect to the `mass' or `kinetic metric' $m_{ij}$. Remarkably, this connection between trajectories and geodesics extends to a system subject to a potential $V$. Indeed, this is the content of Maupertuis' principle of extremization of $\int_{q_1}^{q_2} p dq$ holding energy fixed\cite{lanczos,arnold}. More precisely, the equations of motion (EOM)
	\beq 
	 m_{ki} \ddot x^i(t) = - \pdr_k V - \half \left(m_{ik,j} + m_{jk,i} - m_{ij,k} \right) \dot x^i(t) \: \dot x^j(t)
	 \label{e:Lagrange-eqns-kin-metric-and-V}
	 \eeq
may be regarded as reparametrized geodesic equations for the JM metric,
	\beq
	ds^2 = g_{ij} dx^i dx^j = (E-V) m_{ij} dx^i dx^j
	\eeq
on the classically allowed `Hill' region $E - V \geq 0$. Notice that $\sqrt{2} \int ds = \int p dq = \int (L+E) dt$ so that the length of a geodesic is related to the classical action of the trajectory. The formula for the inverse JM metric $g^{ij} = m^{ij}/(E-V)$ may also be read off from the time-independent Hamilton-Jacobi (HJ) equation $(m^{ij}/2(E-V))\; \pdr_i W \pdr_j W = 1$ by analogy with the rescaled kinetic metric $m^{ij}/2E$ appearing in the free particle HJ equation $(m^{ij}/2E) \pdr_i W \pdr_j W = 1$ (see p.74 of Ref. \cite{Rajeev}). The JM metric is conformal to the kinetic metric and depends parametrically on the conserved energy $E = \half m_{ij} \dot x^i \dot x^j + V$. The geodesic equations 
	\beq
	\label{e:jm-geodesic-equation}
	\ddot x^l(\la) = -  \ov{2}g^{lk}\left(g_{ki,j}+g_{kj,i}-g_{ij,k}\right)\dot x^i(\la) \dot x^j(\la)
	\eeq 
for the JM metric reduce to (\ref{e:Lagrange-eqns-kin-metric-and-V}) under the reparametrisation $d/d\la = (1/\sigma) (d/dt)$ where $\sigma = (E-V)/\sqrt{\cal{T}}$. Here ${\cal T} = \half g_{ij} \dot x^i \dot x^j$ is the conserved `kinetic energy' along geodesics and equals one-half for arc-length parametrization. To obtain $\sigma$, suppose $y^i(t)$ is a trajectory and $z^i(\la)$ the corresponding geodesic. Then at a point $x^i = z^i(\la) = y^i(t)$, the velocities are related by $\sigma \dot z^i = \dot y^i$ leading to
	\beq
	{\cal T} = \half g_{ij} \dot z^i \dot z^j =\fr{E-V}{2}m_{ij} \dot{z}^i\dot{z}^j = \fr{E-V}{2 \sigma^2} m_{ij} \dot{y}^i\dot{y}^j  = \left(\fr{E-V}{ \sigma}\right)^2.
	\eeq
This reparametrization of time may be inconsequential in some cases [e.g. Lagrange rotational solutions where $\sigma$ is a constant since $V$ is constant along the trajectory (see \S \ref{s:stability-tensor})] but may have significant effects in others [e.g. Lagrange homothety solutions where the exponential time-reparametrization regularizes triple collisions (see \S \ref{s:triple-collision-inv-sq-near-collision-geom})] and could even lead to a difference between linear stability of trajectories and corresponding geodesics (see \S \ref{s:stability-tensor}). 

The curvature of the JM metric encodes information on linear stability of geodesics (see \S \ref{s:sectional-curvature-inv-sq-pot}). For example, in the planar isotropic harmonic oscillator with potential $k r^2/2$ in plane polar coordinates, the gaussian curvature $R = 16Ek/(2E-kr^2)^3$ of the JM metric on configuration space is non-negative everywhere indicating stability. In the planar Kepler problem with Hamiltonian $\bfp^2/2m - k/r$, the gaussian curvature of the JM metric $ds^2 = m(E+k/r)(dr^2+r^2d\theta^2)$ is $R =  -{Ek}/ ({m(k+Er)^3})$. $R$ is  everywhere negative/positive for $E$ positive/negative and vanishes identically for $E = 0$. This reflects the divergence of nearby hyperbolic orbits and oscillation of nearby elliptical orbits. Negativity of curvature could lead to chaos, though not always, as the hyperbolic orbits of the Kepler problem show. As noted, chaos could also arise from curvature fluctuations \cite{pettini-2000-phys-rpts}.

\section{Planar three-body problem with inverse-square potential}

\subsection{Jacobi-Maupertuis metric on configuration space and Hopf coordinates}
\label{s:jm-metric-config-space-hopf-coords}

\begin{figure}	
	\centering
	\begin{subfigure}[t]{3in}
		\centering
		\includegraphics[width=3in]{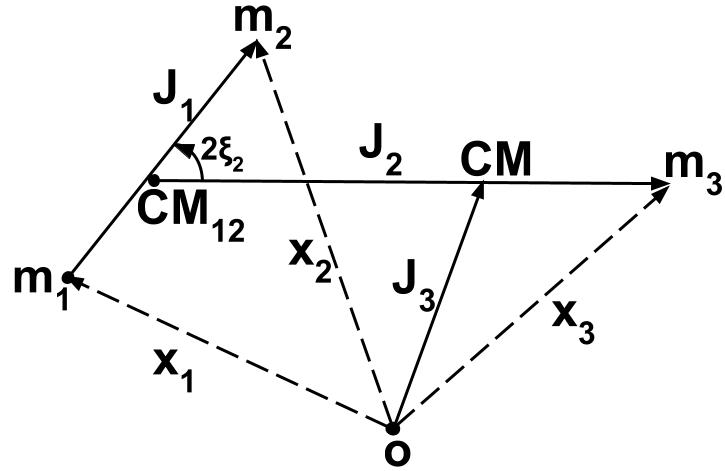}
		\caption{}
		\label{f:jacobi-vectors}		
	\end{subfigure}
	\quad
	\begin{subfigure}[t]{2in}
		\centering
		\includegraphics[width=2in]{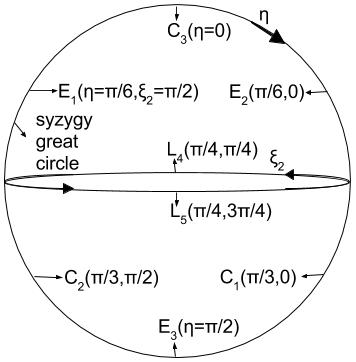}
		\caption{}
		\label{f:shape-sphere-xi2-eta}
	\end{subfigure}
	\quad
	\begin{subfigure}[t]{.9in}
		\centering
		\includegraphics[width=1in]{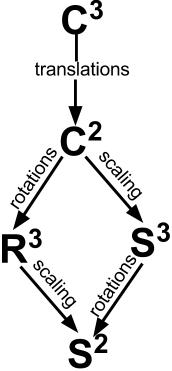}
		\caption{}
		\label{f:flow-chart}
	\end{subfigure}
	\caption{\footnotesize (a) Position vectors $x_{1,2,3}$ of masses relative to origin and Jacobi vectors $J_{1,2,3}$. (b) The shape sphere is topologically a 2-sphere with the three collision points $C_{1,2,3}$ removed, endowed with the quotient JM metric of {\it negative} gaussian curvature. Coordinates and physical locations on the shape sphere are illustrated. 2$\eta$ is the polar angle ($0 \leq \eta \leq \pi/2$). 2$\xi_2$ is the azimuthal angle ($0\leq \xi_2 \leq \pi$). The `great circle' composed of the two longitudes $\xi_2=0$ and $\xi_2=\pi/2$ consists of collinear configurations (syzygies) which include $C_{1,2,3}$ and the Euler points $E_{1,2,3}$. Lagrange points $L_{4,5}$ lie on the equator $\eta = \pi/4$. The shape space $\mR^3$ is a cone on the shape sphere. The origin $r=0$ of shape space is the triple collision point. (c) Flowchart of submersions.}\label{f:shape-sphere-xi2-eta-and-jacobi-vectors}
\end{figure}

We consider the three-body problem  with masses moving on a plane regarded as the complex plane $\bf{C}$. Its $6$D configuration space (with collision points excluded) is identified with ${\mC^3}$. A point on $\mC^3$ represents a triangle on the complex plane with the masses $m_{1,2,3}$ at its vertices $x_{1,2,3} \in \mathbf{C}$. It is convenient to work in Jacobi coordinates (Fig. \ref{f:jacobi-vectors})
	\beq
	\label{e:jacobi-coordinates-on-c3} 
J_1=x_2-x_1, \quad J_2=x_3-\fr{m_1x_1+m_2x_2}{m_1+m_2} \quad \text{and} \quad J_3=\fr{m_1x_1+m_2x_2+m_3x_3}{M_3},
	\eeq
in which the kinetic energy $KE = (1/2) \sum_i m_i |\dot x_i|^2$ remains diagonal:
	\beq
	\label{e:ke-in-jacobi-coordinates-on-c3}
	KE = \half \sum_i M_i | \dot J_i|^2 \;\; \text{where} \;\;
	\ov{M_1}=\ov{m_1}+\ov{m_2}, \;\; \ov{M_2}=\ov{m_3}+\ov{m_1+m_2} \;\;\text{and} \;\; M_3 = \sum_i m_i.
	\eeq
The KE for motion about the center of mass (CM) is $\half (M_1 |\dot J_1|^2 + M_2 |\dot J_2|^2)$. The moment of inertia about the origin $I = \sum_{i=1}^3 m_i |x_i|^2$ too remains diagonal in Jacobi coordinates ($I = \sum_{i=1}^3 M_i |J_i|^2$), while about the CM we have $I_{\rm CM} = M_1 |J_1|^2 + M_2 |J_2|^2$. With $U = -V = \sum_{i < j} G m_i m_j/|x_i - x_j|^2$ denoting the (negative) potential energy, the JM metric for energy $E$ on $\mC^3$ is
	\beqs
	\label{e:jm-metric-in-jacobi-coordinates-on-c3}
	ds^2 = \left( E + U  \right) \sum_{i=1}^3 M_i |dJ_i|^2 \quad \text{where} \;\; U = \fr{G m_1 m_2}{|J_1|^2} + \fr{G m_2 m_3}{|J_2 - \mu_1 J_1|^2} + \fr{G m_3 m_1}{|J_2+\mu_2 J_1|^2}
	\eeqs
and $\mu_i= m_i/(m_1 + m_2)$. Due to the inverse-square potential, $G$ {\it does not} have the usual dimensions. The metric is independent of the CM coordinates $J_3$ and $\bar J_3$, while $J_1,\bar J_1, J_2$ and $\bar J_2$ are invariant under translations $x_i \to x_i + a$ for $a \in \mathbf{C}$. Thus translations act as isometries of (\ref{e:jm-metric-in-jacobi-coordinates-on-c3}). Similarly, we will see that scalings (for $E =0$) and rotations also act as isometries. These isometries also act as symmetries of the Hamiltonian. For instance the dilatation $D = \sum_i \vec x_i \cdot \vec p_i = \sum_i \Re (x_i \bar p_i)$ generates scale transformations $x_i \to \la x_i$ and $p_i \to \la^{-1} p_i$ via Poisson brackets: $\{x_i, D \} = x_i$ and $\{p_i, D \} = - p_i$. Since $\{ H, D \} = - 2 H$, scaling is a symmetry of the Hamiltonian only when energy vanishes.

The study of the geometry of the JM metric is greatly facilitated by first considering the geometry of its quotients by isometries (for instance, geodesics on a quotient lift to horizontal geodesics). Riemannian submersions \cite{oneill} provide a framework to define and compute metrics on these quotients. Suppose $(M,g)$ and $(N,h)$ are two Riemannian manifolds and $f: M \rightarrow N$ a surjection. Then the linearization $df(p): T_p M \to T_{f(p)} N$ is a surjection between tangent spaces. The vertical subspace $V(p) \subseteq T_p M$ is defined to be the kernel of $df$ while its orthogonal complement $\ker(df)^\perp$ with respect to the metric $g$ is the horizontal subspace $H(p)$. $f$ is a Riemannian submersion if it preserves lengths of horizontal vectors, i.e., if the isomorphism $df(p) \colon \ker(df(p))^{\perp} \rightarrow T_{f(p)} N$ is an isometry at each point. The Riemannian submersions we are interested in are associated to quotients of a Riemannian manifold ($M,g$) by the action of a suitable group of isometries $G$. There is a natural surjection $f$ from $M$ to the quotient $M/G$. Requiring $f$ to be a Riemannian submersion defines the quotient metric on $M/G$: the inner product of a pair of tangent vectors $(u,v)$ to $M/G$ is defined as the inner product of {\it any} pair of horizontal preimages under the map $df$.

The surjection $\left( J_1,\bar J_1, J_2,\bar J_2, J_3,\bar J_3\right) \mapsto\left( J_1,\bar J_1, J_2,\bar J_2\right)$ defines a submersion from configuration space $\mC^3$ to its quotient $\mC^2$ by translations. The vertical and horizontal subspaces are spanned by $\pdr_{J_3}, \pdr_{\bar J_3}$ and $\pdr_{ J_1}, \pdr_{\bar J_1}, \pdr_{J_2}, \pdr_{\bar J_2}$ respectively. Requiring the submersion to be Riemannian, the quotient metric on $\mC^2$ is
	\beq
	\label{e:jm-metric-in-jacobi-coordinates-on-c2}
	ds^2=( E + U ) (M_1 \; |d J_1|^2 + M_2 \; |d J_2|^2).
	\eeq
It is convenient to define rescaled coordinates on $\mC^2$, $z_i = \sqrt{M_i} \:  J_i$, in terms of which (\ref{e:jm-metric-in-jacobi-coordinates-on-c2}) becomes $ds^2 = (E + U) (|dz_1|^2 + |dz_2|^2)$. The kinetic energy in the CM frame is $KE = (1/2) (|\dot z_1|^2 + |\dot z_2|^2 )$ while $I_{\rm CM} = |z_1|^2 + |z_2|^2$.

We now specialize to equal masses ($m_i = m$) so that $M_1 = m/2$,$M_2={2m}/{3}$ and $\mu_i = 1/2$. The metric on $\mC^2$ is seen to be conformal to the flat Euclidean metric  via the conformal factor $E + U$:
	\beq
	\label{e:jm-metric-in-rescaled-jacobi-coordinates-for-eq-mass-on-c2}
	ds^2 = \left(E+\fr{G m^3}{2|z_1|^2}+\fr{2 G m^3}{3|z_2-\ov{\sqrt{3}}z_1|^2}+\fr{2 G m^3}{3|z_2+\ov{\sqrt{3}}z_1|^2}\right) \left(|dz_1|^2+|dz_2|^2 \right) .
	\eeq
Rotations U$(1)$ act as a group of isometries of $\mC^2$, taking $\left(z_1,z_2\right) \mapsto\left(e^{i \tht}z_1,e^{i \tht}z_2\right)$ and leaving the conformal factor invariant. Moreover if $E = 0$, then scaling $z_i \mapsto \la z_i$ for $\la \in {\bf R}^+$ is also an isometry. Thus we may quotient the configuration manifold $\mC^2$ successively by its isometries. We will see that $\mC^2/$U$(1)$ is the shape space $\mR^3$ and $\mC^2$/scaling is $\mS^3$. Furthermore the quotient of $\mC^2$ by both scaling and rotations leads to the shape sphere $\mS^2$ (see Fig. \ref{f:flow-chart}, note that collision points are excluded from $\mC^2, \mR^3, \mS^3$ and $\mS^2$). Points on shape space $\mR^3$ represent oriented congruence classes of  triangles while those on the shape sphere $\mS^2$ represent oriented similarity classes of triangles. Each of these quotient spaces may be given a JM metric by requiring the projection maps to be Riemannian submersions. The geodesic dynamics on $\mC^2$ is clarified by studying its projections to these quotient manifolds. We will now describe these Riemannian submersions explicitly in local coordinates. This is greatly facilitated by choosing coordinates (unlike $z_1, z_2$) on $\mC^2$ in which the KVFs corresponding to the isometries point along coordinate vector fields. As we will see, this ensures that the vertical subspaces in the associated Riemannian submersions are spanned by coordinate vector fields. Thus we introduce the Hopf coordinates $(r, \eta, \xi_1, \xi_2)$ on $\mC^2$ via the transformation
	\beq
	z_1=r e^{i ( \xi_1+ \xi_2)} \sin\eta \quad \text{and} \quad z_2=r e^{i ( \xi_1- \xi_2)} \cos\eta.
	\label{e:Hopf-coords}
	\eeq
Here the radial coordinate $r = \sqrt{|z_1|^2 + |z_2|^2} = \sqrt{I_{\rm CM}} \geq 0$ is a measure of the size of the triangle with masses at its vertices. $\xi_2$ determines the relative orientation of $z_1$ and $z_2$ while $\xi_1$ fixes the orientation of the triangle as a whole. More precisely, $2 \xi_2$ is the angle from the rescaled Jacobi vector $z_2$ to $z_1$ while $2 \xi_1$ is the sum of the angles subtended by $z_1$ and $z_2$ with the horizontal axis in Fig \ref{f:jacobi-vectors}. Thus we may take $0 \leq \xi_1 + \xi_2 \leq 2\pi$ and $0 \leq \xi_1 -\xi_2 \le 2 \pi$ or equivalently, $-\pi\leq\xi_2\leq\pi$ and $|\xi_2| \leq \xi_1 \leq 2\pi-|\xi_2|$. Finally, $0 \leq \eta \leq \pi/2$ measures the relative magnitudes of $z_1$ and $z_2$, indeed $\tan \eta = |z_1|/|z_2|$. When $r$ is held fixed, $\eta, \xi_1$ and $\xi_2$ furnish the standard Hopf coordinates parametrizing the three sphere $|z_1|^2 + |z_2|^2 = r^2$. For fixed $r$ and $\eta$, $\xi_1 + \xi_2$ and $\xi_1 - \xi_2$ are periodic coordinates on tori. These tori foliate the above three-sphere as $\eta$ ranges between $0$ and $\pi/2$. Furthermore, as shown in \S \ref{s:quotient-metrics}, $2 \eta$ and $2 \xi_2$ are polar and azimuthal angles on the two-sphere obtained as the quotient of $\mS^3$ by rotations via the Hopf map.

Let us briefly motivate these coordinates and the identification of the above quotient spaces. We begin by noting that the JM metric (\ref{e:jm-metric-in-rescaled-jacobi-coordinates-for-eq-mass-on-c2}) on $\mC^2$ is conformal to the flat Euclidean metric $|dz_1|^2 + |dz_2|^2$. Recall that the cone on a Riemannian manifold $(M, ds^2_M)$ is the Cartesian product $\mathbf{R}^+ \times M$ with metric $dr^2 + r^2 ds_M^2$ where $r > 0$ parameterizes $\mathbf{R}^+$. In particular, Euclidean $\mathbf{C}^2$ may be viewed as a cone on the round three sphere $\mathbf{S}^3$. If $\bf{S}^3$ is parameterized by Hopf coordinates $\eta, \xi_1$ and $\xi_2$, then this cone structure allows us to use $r, \eta, \xi_1$ and $\xi_2$ as coordinates on $\bf{C}^2$. Moreover, the Hopf map defines a Riemannian submersion from the round $\mathbf{S}^3$ to the round two sphere $\mathbf{S}^2$ \footnote{ The Hopf map $\mathbf{S}^3 \to \mathbf{S}^2$ is often expressed in Cartesian coordinates. If  $|z_1|^2 + |z_2|^2 = 1$ defines the unit-$\mathbf{S}^3 \subset \mathbf{C}^2$ and $w_1^2 + w_2^2 + w_3^2 = 1/4$ defines a 2-sphere of radius $1/2$ in $\mathbf{R}^3$, then $w_3 = \left(|z_2|^2-|z_1|^2\right)/2$ and $w_1+ i w_2 = z_1 \bar z_2$. Using Eq.\ref{e:Hopf-coords}, we may express the Cartesian coordinates $w_i$ in terms of Hopf coordinates:
	\beq \nonumber
	2 w_3 = {r^2}\cos2\eta,
\quad 2 w_1 = r^2 \sin(2\eta)\cos(2\xi_2)\quad \text{and} \quad 2 w_2 = r^2 \sin(2\eta)\sin(2\xi_2).
	\eeq
}. Indeed, if we use Hopf coordinates $\eta, \xi_1, \xi_2$ on $\bf{S}^3$, then the Hopf map takes $(\eta, \xi_1, \xi_2) \mapsto (\eta, \xi_2) \in \bf{S}^2$. In general, if $M \to N$ is a Riemannian submersion, then there is a natural submersion from the cone on $M$ to the cone on $N$ \footnote{Let $f:(M,g)\mapsto(N,h)$ be a Riemannian submersion with local coordinates $m^i$ and $n^j$. Let $(r, m^i)$ and $(r,n^j)$ be local coordinates on the cones $C(M)$ and $C(N)$. Then $\tl f: (r, m) \mapsto (r, n)$ defines a submersion from $C(M)$ to $C(N)$. Consider a horizontal vector $a \pdr_r + b_i \pdr_{m_i}$ in $T_{(r,m)}C(M)$. We will show that $d\tl f$ preserves its length. Now, if $df(b_i \pdr_{m_i}) = c_i \pdr_{n_i}$ then $d\tilde f(a \pdr_r + b_i \pdr_{m_i}) = a \pdr_r + c_i \pdr_{n_i}$. Since $\pdr_r \perp \pdr_{m^i}$, $||a \pdr_r+ b_i \pdr_{m_i}||^2 = a^2 + r^2  \|b_i \pdr_{m_i}\|^2$ $= a^2 + r^2\|c_i \pdr_{n_i}\|^2$ as $f$ is a Riemannian submersion. Moreover $a^2 + r^2 \|c_i \pdr_{n_i}\|^2 = \| a \pdr_r + c_i \pdr_{n_i} \|^2$ since $\pdr_r \perp \pdr_{n^i}$. Thus $\tilde f$ is a Riemannian submersion.
}. In particular, the Hopf map extends to a Riemannian submersion from the cone on the round $\bf{S}^3$ to the cone on the round $\bf{S}^2$, i.e. from Euclidean $\bf{C}^2$ to Euclidean $\bf{R}^3$ taking $(r, \eta, \xi_1, \xi_2) \mapsto (r, \eta, \xi_2)$. As the conformal factor is independent of rotations, the same map defines a Riemannian submersion from $\mC^2$ with the JM metric to shape space $\mR^3$ with its quotient JM metric. Finally, for $E=0$, scaling ${\vec r} \to \la {\vec r}$ defines an isometry of the quotient JM metric on shape space $\mR^3$. Quotienting by this isometry we arrive at the shape sphere $\mS^2$ with Montgomery's `pair of pants' metric. Alternatively, we may quotient $\mC^2$ first by the scaling isometry of its JM metric to get $\mS^3$ and then by rotations to get $\mS^2$ (see Fig. \ref{f:flow-chart}).

With these motivations, we express the equal-mass JM metric on $\mC^2$ in Hopf  coordinates [generalization to unequal masses is obtained by replacing $Gm^3 h$ below with $\tl h(\eta, \xi_2)$ given in Eq. (\ref{e:unequal-mass-htilde})]:
	\beqs
	\label{e:c2-metric-in-r-eta-xi1-xi2-coordinates}
	ds^2 = \left( E + \fr{Gm^3 h(\eta,\xi_2)}{r^2} \right) \left(dr^2+r^2\left(d\eta^2+ d\xi_1^2-2\cos2\eta\;d\xi_1\;d\xi_2+ d\xi_2^2\right)\right).
	\eeqs
It is convenient to write $h(\eta,\xi_2) = v_1+ v_2 + v_3$ where $v_1 = r^2/(m |x_2 - x_3|^2)$ is proportional to the pairwise potential between $m_2$ and $m_3$ and cyclic permutations thereof. The $v_i$ are rotation and scale-invariant, and therefore functions only of $\eta$ and $\xi_2$ in Hopf coordinates:
	\beq
	\label{e:conformal-prefactor-h}
	v_{1,2} = \fr{2}{\left( 2 + \cos2\eta \mp \sqrt{3}   \sin2\eta\; \cos2\xi_2\right)} \quad \text{and} \quad v_3 = \fr{1}{2\sin^2\eta}.
	\eeq 
Notice that $h \to \infty$ at pairwise collisions. The $v_i$'s have the common range $1/2 \leq v_i < \infty$ with $v_3 = 1/2$ when $m_3$ is at the CM of $m_1$ and $m_2$ etc. We also have $h \geq 3$ with equality when $v_1 = v_2 = v_3$, corresponding to Lagrange configurations with masses at vertices of an equilateral triangle. To see this, we compute the moment of inertia $I_{\rm CM}$ in two ways. On the one hand $I_{\rm CM}=|z_1|^2+|z_2|^2=r^2$ . On the other hand, for equal masses the CM lies at the centroid of the triangle defined by masses. Thus $I_{\rm CM}$ is $(4m/9) \times$ the sum of the squares of the medians, which by Apollonius' theorem is equal to $(3/4) \times$ the sum of the squares of the sides. Hence $I_{\rm CM}=\sum_{i=1}^3 r^2/3 v_i$. Comparing, we get  $\sum_{i=1}^3 1/v_i = 3$. Since the arithmetic mean is bounded below by the harmonic mean,
	\beq
	\label{ineq:conformal-factor-h}
	{h}/{3} = {\left( v_1 + v_2 + v_3 \right)}/{3} \geq 3 \left( {v_1}^{-1} + {v_2}^{-1} + {v_3}^{-1} \right)^{-1} = 1.
	\eeq

{\fl \bf Lagrange, Euler, collinear and collision configurations:}
The geometry of the JM metric displays interesting behavior at Lagrange and collision configurations on $\mC^2$ and its quotients. We identify their locations in Hopf coordinates for {\it equal} masses. The Jacobi vectors in Hopf coordinates are 
	\beq
	J_1=\sqrt{\fr{2}{m}} r e^{i ( \xi_1+ \xi_2)} \sin\eta \quad \text{and} \quad 
	J_2=\sqrt{\fr{3}{2 m}}r e^{i ( \xi_1- \xi_2)} \cos\eta.
	\eeq
At a Lagrange configuration, $m_{1,2,3}$ are at vertices of an equilateral triangle. So $|J_2| = \sqrt{3}|J_1|/2$ (i.e. $\eta = \pi/4$) and $J_2$ is $\perp$ to $J_1$ (i.e. $\xi_2 = \pm \pi/4$, the sign being fixed by the orientation of the triangle). So Lagrange configurations $L_{4,5}$ on $\mC^2$ occur when $\eta = \pi/4$ and $\xi_2=\pm\pi/4$ with $r$ and $\xi_1$ arbitrary. On quotients of $\mC^2$, $L_{4,5}$ occur at the images under the corresponding projections. Since $2 \eta$ and $2 \xi_2$ are polar and azimuthal angles on the shape sphere, $L_{4,5}$ are at diametrically opposite equatorial locations (see Fig. \ref{f:shape-sphere-xi2-eta}). Collinear configurations (syzygies) occur when $J_1$ and $J_2$ are (anti)parallel, i.e. when $\xi_2 = 0$ or $\pi/2$, with other coordinates arbitrary. On the shape sphere, syzygies occur on the `great circle' through the poles corresponding to the longitudes $2\xi_2 = 0$ and $\pi$. Collisions are special collinear configurations. By $C_i$ we denote a collision of particles other than the $i^{\rm th}$ one. So $C_3$ corresponds to $J_1=0$ which lies at the `north pole' ($\eta=0$) on $\mS^2$. $m_2$ and $m_3$ collide when $J_2 =  J_1/2$ so  $\eta = \pi/3$ and $\xi_2 = 0$ at $C_1$. Similarly, at $C_2$, $J_2 = - J_1/2$ which corresponds to $\eta = \pi/3$ and $\xi_2=\pi/2$. The Euler configurations $E_i$ for equal masses are collinear configurations where mass $m_i$ is at the midpoint of the other two. 

Finally, we note that the azimuth and co-latitude ($\tht$ and $\phi$) \cite{montgomery-pants} are often used as coordinates on the shape sphere, so that $L_{4,5}$ are at the poles while $C_{1,2,3}$ and $E_{1,2,3}$ lie on the equator. This coordinate system makes the symmetry under permutations of masses explicit, but is not convenient near any of the collisions (e.g. sectional curvatures can be discontinuous). On the other hand, our coordinates $\eta$ and $\xi_2$, which are related to $\tht$ and $\phi$ by suitable rotations,
	\beq \nonumber
	\label{e:thtphi-xieta}
	\sin\phi =  \cos(2\eta-\pi /2)\sin(2\xi_2), \;\; \cos\phi \sin\tht=\cos(2\eta-\pi/2)\cos(2\xi_2), 
\;\; \cos\phi \cos\tht=\sin(2\eta-\fr{\pi}{2}),
	\eeq
are convenient near $C_3$ but not near $E_3$ or $C_{1,2}$ (sectional curvatues can be discontinuous, see \S \ref{s:sectional-curvature-inv-sq-pot}). The neighborhoods of the latter configurations may be studied by re-ordering the masses.

\subsection{Quotient JM metrics on shape space, the three-sphere and the shape sphere}
\label{s:quotient-metrics}

{\fl \bf Submersion from $\mC^2$ to shape space $\mR^3$:}
Rotations $z_j \to e^{i \tht} z_j$ act as isometries of the JM metric (\ref{e:c2-metric-in-r-eta-xi1-xi2-coordinates}) on $\mC^2$. In the Hopf coordinates of Eq. (\ref{e:Hopf-coords}),
	\beq
	z_1=r e^{i ( \xi_1+ \xi_2)} \sin\eta \quad \text{and} \quad z_2=r e^{i ( \xi_1- \xi_2)} \cos\eta, \quad 
	\eeq
rotations are generated by translations  $\xi_1 \to \xi_1 + \tht$ and a discrete shift $\xi_2 \to \xi_2 + \pi$ (${\rm mod} \; 2\pi$). The shift in $\xi_2$ rotates $z_i \mapsto -z_i$, which is not achievable by a translation in $\xi_1$ due to its restricted range, $|\xi_2| \leq \xi_1 \leq 2\pi - |\xi_2|$ and $-\pi \le \xi_2 \le \pi$. To quotient by this isometry, we define a submersion from $\mC^2 \to \mR^3$ taking 
	\beq
	(r, \eta, \xi_1,\xi_2) \mapsto (r, \eta, \xi_2)\quad \text{if} \quad \xi_2 \ge 0 \quad \text{and} \quad  (r,\eta, \xi_1,\xi_2) \mapsto (r,\eta, \xi_2+ \pi) \quad \text{if} \quad \xi_2 < 0.
	\eeq
The radial, polar and azimuthal coordinates on $\mR^3$ are given by	$r$, $2\eta$ and $2\xi_2$ with $m_1$-$m_2$ collisions occurring on the ray $\eta =0$. Under the linearization of this submersion at a point $p \in \mC^2$, $V(p)$ is spanned by $\pdr_{\xi_1}$ and $H(p)$ by $\pdr_r$, $\pdr_\eta$ and $\cos 2 \eta \; \pdr_{\xi_1} + \pdr_{\xi_2}$. These horizontal basis vectors are mapped respectively to $\pdr_r$, $\pdr_\eta$ and $\pdr_{\xi_2}$ under the linearization of the map. Requiring lengths of horizontal vectors to be preserved we arrive at the following quotient JM metric on $\mR^3$, conformal to the flat metric on $\mR^3$:
	\beq
	\label{e:r3-metric-in-r-eta-xi2-coordinates}
	ds^2=  \left( E + \fr{Gm^3 h(\eta,\xi_2)}{r^2} \right)\left(dr^2+r^2\left(d\eta^2+\sin^2 2\eta \;d\xi_2^2\right)\right).
	\eeq
This metric may also be viewed as conformal to a cone on a round $2$-sphere of radius one-half, since $0 \le 2 \eta \le \pi$ and $0 \le 2 \xi_2 \le 2 \pi$ are the polar and azimuthal angles.

{\fl \bf Submersion from shape space to the shape sphere:}
The group $\bf{R}^+$ of scalings $(r, \eta, \xi_2) \mapsto (\la r, \eta, \xi_2)$ acts as an isometry of the {\it zero-energy} JM metric (\ref{e:r3-metric-in-r-eta-xi2-coordinates}) on shape space $\mR^3$. The orbits are radial rays emanating from the origin (and the triple collision point at the origin, which we exclude). The quotient space $\mR^3/{\rm scaling}$ is the shape sphere $\mS^2$. We define a submersion from shape space to the shape sphere taking $(r, \eta, \xi_2) \mapsto (\eta, \xi_2)$. Under the linearization of this map at $p \in \mR^3$, $V(p) = \text{span}(\pdr_r)$. Its orthogonal complement $H(p)$ is spanned by $\pdr_\eta$ and $\pdr_{\xi_2}$ which project to $\pdr_\eta$ and $\pdr_{\xi_2}$ on $\mS^2$. Requiring the submersion to be Riemannian, we get the quotient `pair of pants' JM metric on the shape sphere which is conformal to the round metric on a $2$-sphere of radius one-half:
	\beq
	\label{e:s2-metric-in-eta-xi2-coordinates}
	ds^2 = Gm^3 h(\eta,\xi_2) \left(d\eta^2+\sin^2 2\eta \;d\xi_2^2\right).
	\eeq

{\fl \bf Submersion from $\mC^2$ to $\mS^3$ and then to $\mS^2$:}
For zero energy, it is also possible to quotient the JM metric (\ref{e:c2-metric-in-r-eta-xi1-xi2-coordinates}) on $\mC^2$, first by its scaling isometries to get $\mS^3$ and then by rotations to arrive at the shape sphere. Interestingly, it follows from the Lagrange-Jacobi identity that when $E$ and $\dot I$ vanish, $r$ is constant and the motion is confined to a $3$-sphere embedded in $\mC^2$. To quotient by the scaling isometries $(r, \eta, \xi_1, \xi_2) \mapsto (\la r, \eta, \xi_1, \xi_2)$ of $\mC^2$, we define the submersion $(r, \eta, \xi_1, \xi_2) \mapsto (\eta, \xi_1, \xi_2)$ to $\mS^3$, with ranges of coordinates as on $\mC^2$. The vertical subspace is spanned by $\pdr_r$ while $\pdr_\eta$, $\pdr_{\xi_1}$ and $\pdr_{\xi_2}$ span the horizontal subspace. The latter are mapped to $\pdr_\eta$, $\pdr_{\xi_1}$ and $\pdr_{\xi_2}$ on $\mS^3$. The submersion is Riemannian provided we endow $\mS^3$ with the following conformally-round metric 
	\beq
	\label{e:s3-metric-in-eta-xi1-xi2-coordinates}
ds^2=Gm^3 h\left(\eta,\xi_2\right)\left(d\eta^2+ d\xi_1^2-2\cos2\eta\;d\xi_1\;d\xi_2+ d\xi_2^2\right). 
	\eeq
Rotations generated by $\xi_1 \to \xi_1 + \tht$ and $\xi_2 \to \xi_2 + \pi$ (mod $2\pi$) act as isometries of this metric on $\mS^3$. We quotient by rotations to get the metric (\ref{e:s2-metric-in-eta-xi2-coordinates}) on $\mS^2$ via the Riemannian submersion defined by 
	\beq
	(\eta,\xi_1,\xi_2) \mapsto (\eta,\xi_2) \quad \text{if} \quad \xi_2 \ge 0 \quad \text{and} \quad (\eta,\xi_1,\xi_2) \mapsto (\eta,\xi_2+\pi) \quad \text{if} \quad \xi_2 < 0. 
	\eeq
\subsection{JM metric in the near-collision limit and its completeness}
\label{s:geodesic-completeness}

The equal-mass JM metric components on configuration space $\mC^2$ and its quotients blow up at $2$- and $3$-body collisions. However, we study the geometry in the neighbourhood of collision configurations and show that the curvature remains finite in the limit. Remarkably, it takes infinite geodesic time for collisions to occur which we show by establishing the geodesic completeness of the JM metric on $\mC^2$ and its quotients. By contrast, collisions can occur in finite time for the Newtonian $3$-body evolution. The JM geodesic flow avoids finite time collisions by reparametrizing time along Newtonian trajectories (see Eq. \ref{e:jm-geodesic-equation}). Thus the geodesic reformulation of the inverse-square $3$-body problem `regularizes' pairwise and triple collisions.

\subsubsection{Geometry near pairwise collisions}
\label{s:near-pairwise-collision-geometry}

For equal masses (see \S\ref{s:jm-metric-config-space-hopf-coords}), the first pair of masses collide when $\eta=0$ (with other coordinates arbitrary) while the other two binary collisions occur at $C_1$ and $C_2$ (see Fig. \ref{f:shape-sphere-xi2-eta}). Triple collisions occur when $r = 0$. Unlike for the Newtonian potential, sectional curvatures on coordinate $2$-planes are finite at pairwise and triple collisions, though some JM metric (\ref{e:c2-metric-in-r-eta-xi1-xi2-coordinates}) and Riemann tensor components blow up. It is therefore interesting to study the near-collision geometry of the JM metric.

The geometry of the equal-mass JM metric in the neigbourhood of a binary collision is the same irrespective of which pair of bodies collide. Since Hopf coordinates are particularly convenient around $\eta = 0$, we focus on collisions between the first pair of masses. Montgomery (see eqn. 3.10c of \cite{montgomery-pants}) studied the near-collision geometry on $\mS^2$ and showed that it is geodesically complete. Let us briefly recall the argument. Expanding the equal-mass $\mS^2$ metric (\ref{e:s2-metric-in-eta-xi2-coordinates}) around the collision point $\eta = 0$, we get
	\beq
	ds^2 \approx \left(\fr{G m^3}{2\eta^2}\right) \left(d\eta^2+4\eta^2 \;d\xi_2^2\right) = \frac{G m^3}{2\rho^2}(d\rho^2 + \rho^2 d\chi^2)
	\label{e:S2-near-collision-metric-inv-sq}
	\eeq
where $\rho = 2 \eta$ and $\chi = 2 \xi_2$. $\pdr_{\chi}$ is a KVF, so `radial' curves with constant $\chi$ are geodesics. Approaching $\rho = 0$ along a `radial' geodesic shows that the collision point $\rho = 0$ is at an infinite distance $(\sqrt{G m^3/2} \int_{\rho_0}^0 d\rho/\rho)$ from any point $(\rho_0,\chi)$ in its neighborhood $(0< \rho_0 \ll 1)$. The symmetry of the metric under exchange of masses ensures that the same holds for the other two collision points: geodesics may be extended indefinitely. Thus the shape sphere ($\mS^2$ with three collision points excluded) is geodesically complete. To clarify the near-collision geometry let $d \la = -d\rho/\sqrt{2}\rho$ or $\la = - \log(\rho/\rho_0)/\sqrt{2}$. This effectively stretches out the neighborhood of the collision point $\la = \infty$. The asymptotic metric $ds^2 = Gm^3 \left( d \la^2 + d\chi^2/2 \right)$ for $0 \leq \chi \leq 2\pi$ and $\la \ge 0$ is the metric on a semi-infinite right-circular cylinder of radius $\sqrt{Gm^3/2}$ with $\la$ the coordinate along the height and $\chi$ the azimuthal angle. Thus the JM metric looks like that of a semi-infinite cylinder near any of the collision points. 

More generally, for {\it unequal} masses, the near-collision metric (\ref{e:S2-near-collision-metric-inv-sq}) is $ds^2 \approx \fr{G m_1 m_2 M_1}{2\eta^2} \left(d\eta^2 + 4\eta^2 d\xi_2^2\right)$  (see Eq. (\ref{e:jm-metric-in-jacobi-coordinates-on-c3}-\ref{e:Hopf-coords})) and essentially the same argument implies that the JM metric on the shape sphere is geodesically complete for arbitrary masses.

Since $\mS^2$ arises as a Riemannian submersion of $\mR^3$, $\mS^3$ and $\mC^2$, the infinite distance to binary collision points on the shape sphere can be used to show that the same holds on each of the higher dimensional manifolds. To see this, consider the submersion from (say) $\mC^2$ to $\mS^2$. Any curve $\tl \g$ on $\mC^2$ maps to a curve $\g$ on $\mS^2$ with $l(\tl \g) \geq l(\g)$ since the lengths of horizontal vectors are preserved. If there was a binary collision point at finite distance on $\mC^2$, there would have to be a geodesic of finite length ending at it. However, such a geodesic would project to a curve on the shape sphere of finite length ending at a collision point, contradicting its completeness. 

Thus we have shown that the JM metrics (necessarily of zero energy) on $\mS^2$ and $\mS^3$ with binary collision points removed, are geodesically complete for arbitrary masses. On the other hand, to examine completeness on $\mC^2$ and $\mR^3$ we must allow for triple collisions as well as non-zero energy. Geodesic completeness in these cases is shown in \S\ref{s:triple-collision-inv-sq-near-collision-geom}. In the sequel we examine the near-collision geometry on $\mR^3$, $\mS^3$ and $\mC^2$ in somewhat greater detail by Laurent expanding the JM metric components around $\eta = 0$ and keeping only leading terms.

{\fl \bf Shape space geometry near binary collisions:} The equal-mass shape space metric around $\eta=0$, in the leading order, becomes
	\beq
	ds^2 \approx \fr{G m^3}{2\eta^2 r^2} \left( dr^2+ r^2  \left(d\eta^2+4\eta^2 \;d\xi_2^2\right)\right) = G m^3 \left( \fr{2 dr^2}{\rho^2 r^2} + \frac{d\rho^2}{2\rho^2}  +\frac{d\chi^2}{2} \right),
	\eeq
where $\rho=2\eta$ and $\chi=2\xi_2$. We define new coordinates $\la$ and $\kappa$ by $d \la = -d\rho/\sqrt{2}\rho$, $d\kappa= dr/r$ so that $\rho = \rho_0 e^{-\sqrt{2}\la}$.  In these coordinates the collision occurs at $\la = \infty$. The asymptotic metric is
	\beq
	ds^2 \approx G m^3 \left( \fr{2}{\rho_0^2}e^{2 \sqrt{2}\la} d\kappa^2 + d \la^2 + \half d\chi^2\right)
	\eeq
where $0 \leq \chi \leq 2 \pi$ (periodic), $\la \geq 0$ and $-\infty < \kappa < \infty$. This metric has a constant scalar curvature of $-4/Gm^3$. The sectional curvature in the $\pdr_\la-\pdr_\kappa$ plane is equal to $-2/Gm^3$, it vanishes in the other two coordinate planes. These values of scalar and sectional curvatures agree with the limiting values at the $1$-$2$ collision point calculated for the full metric on shape space. The near-collision topology of shape space is that of the product manifold $\bf{S}^1_\chi \times \bf{R}^+_\la \times \bf{R}_\kappa$.

{\fl \bf Near-collision geometry on $\mC^2$:} The equal-mass JM metric in leading order around $\eta = 0$ is
	\beq
	ds^2 \approx \fr{G m^3}{2\eta^2 r^2} \left( dr^2+ r^2  \left(d\eta^2+d\xi_1^2-2(1-2\eta^2) d\xi_1 d\xi_2+d\xi_2^2\right)\right).
	\eeq
Let us define new coordinates $\la,\kappa,\xi_\pm$ such that $d \la = -d\eta/\sqrt{2}\eta$, $d\kappa= -dr/r$ and $\xi_\pm=\xi_1\pm\xi_2$. $0 \leq \xi_\pm \leq 2\pi$ are periodic coordinates parametrizing a torus. The asymptotic metric is
	\beq
	ds^2 \approx G m^3\left( \fr{d\kappa^2}{2\eta^2}  + d \la^2 + \ov{2\eta^2} d\xi_-^2 + \ov{2} d\xi_+^2\right)
	\label{e:C2-near-collision-metric}
	\eeq
where $\eta = \eta_0 e^{-\sqrt{2}\la}$.  This metric has a constant scalar curvature $-12/Gm^3$. The sectional curvature of any coordinate plane containing $\pdr_{\xi_+}$ vanishes due to the product form of the metric. The sectional curvatures of the remaining coordinate planes ($\pdr_\kappa-\pdr_\la, \pdr_\kappa - \pdr_{\xi_-}, \pdr_{\xi_-}-\pdr_\la$) are equal to $-2/Gm^3$. The scalar and sectional curvatures (of corresponding planes) of this metric agree with the limiting values computed from the full metric on $\mC^2$. 

{\fl \bf Near-collision geometry on $\mS^3$:} The submersion $\mC^2 \to \mS^3$ takes $(\kappa, \la, \xi_\pm) \mapsto ( \la, \xi_\pm)$. As the coordinate vector fields on $\mC^2$ are orthogonal, from (\ref{e:C2-near-collision-metric}) the asymptotic metric on $\mS^3$ near the $1$-$2$ collision point is 
	\beq
	ds^2 \approx G m^3\left( d \la^2 + \ov{2\eta^2} d\xi_-^2 + \ov{2} d\xi_+^2\right).
	\eeq
This metric has a constant scalar curvature equal to $-4/Gm^3$. The sectional curvatures on the $\la- \xi_-$ coordinate 2-plane is $-2/Gm^3$ while it vanishes on the other two coordinate 2-planes.

\subsubsection{Geometry on  $\mR^3$ and $\mC^2$ near triple collisions}

\label{s:triple-collision-inv-sq-near-collision-geom}

We argue that the triple collision configuration (which occurs at $r=0$ on $\mC^2$ or shape space $\mR^3$) is at infinite distance from other configurations with respect to the equal-mass JM metrics (Eqs. (\ref{e:c2-metric-in-r-eta-xi1-xi2-coordinates}),(\ref{e:r3-metric-in-r-eta-xi2-coordinates})), which may be written in the form:
	\beq
	ds^2 = (G m^3 h/r^2) dr^2 + G m^3 \: h \: g_{i j} \: dx^i dx^j.
	\eeq
$g_{ij}$ is the positive (round) metric on $\mathbf{S}^3$ ($x^i = (\eta, \xi_1, \xi_2)$) or $\mathbf{S}^2$ ($x^i = (\eta, \xi_2)$) of radius one-half:
	\beq
	g^{\mC^2}_{ij} = \colvec{3}{1 & 0 & 0}{0 & 1 & -\cos 2\eta}{0 & -\cos 2\eta & 1}\quad \text{and} \quad g^{\mR^3}_{ij} =  \colvec{2}{1 & 0 }{0 & \sin 2\eta}.
	\label{e:kinetic-metric-s3-s2}
	\eeq
Together with our results on pairwise collisions (\S \ref{s:near-pairwise-collision-geometry}), it will follow that the manifolds are geodesically complete. As a consequence, the geodesic flow reformulation of the $3$-body problem regularizes triple collisions. To show that triple collision points are at infinite distance we will use the previously obtained lower bound on the conformal factor, $h(\xi_2, \eta) \geq 3$ (see Eqn. \ref{ineq:conformal-factor-h}).

Let $\g(t)$ be a curve joining a non-collision point $\g(t_0) \equiv (r_0, x^i_0)$ and the triple collision point $\g(t_1) \equiv (r=0, x^i_1)$. We show that its length $l(\g)$ is infinite. Since $G m^3 h g_{ij}$ is a positive matrix,
	\beq
	l(\g) = \int_{t_0}^{t_1} dt\sqrt{\fr{G m^3 h}{r^2} \dot r^2 + G m^3 h g_{i j} \dot x^i \dot x^j} \;  \geq \; \int_{t_0}^{t_1} dt  \sqrt{\fr{G m^3 h}{r^2} \dot r^2 }.
	\eeq
Now using $|\dot r| \geq - \dot r$ and $h \geq 3$, we get
	\beq
	l(\g) \geq -\sqrt{3 G m^3} \int_{t_0}^{t_1}  \fr{\dot r}{ r} dt = \sqrt{3 G m^3}\int_{0}^{r_0}  \fr{d r}{ r} = \infty .
	\eeq	
In particular, a geodesic from a non-collision point to the triple collision point has infinite length. Despite appearances, the above inequality $l(\g) \geq \sqrt{3 G m^3} \int_0^{r_0} dr/r$ does not imply that radial curves are always geodesics. This is essentially because $h$ along $\g$ may be less than that on the corresponding radial curve. However, if $(\eta, \xi_1, \xi_2)$ is an angular location where $h$ is minimal (locally), then the radial curve with those angular coordinates is indeed a geodesic because a small perturbation to the radial curve increases $h$ and consequently its length. The global minima of $h$ ($h = 3$) occur at the Lagrange configurations $L_{4,5}$ and local minima ($h = 9/2$) are at the Euler configurations $E_{1,2,3}$ indicating that radial curves at these angular locations are geodesics. In fact, the Christoffel symbols $\G^i_{rr}$ vanish for $i = \eta, \xi_1, \xi_2$ at $L_{4,5}$ and at $E_{1,2,3}$ so that radial curves $\g = (r(t), x^i_0)$ satisfying $\ddot r + \G^r_{rr} \dot r^2= 0$ are geodesics.
  
These radial geodesics at minima of $h$ describe Lagrange and Euler homotheties (where the masses move radially inwards/outwards to/from their CM which is the center of similitude). These homotheties take infinite (geodesic) time to reach the triple collision. By contrast, the corresponding Lagrange and Euler homothety solutions to Newton's equations reach the collision point in finite time. This difference is due to an exponential time-reparametrization of geodesics relative to trajectories. In fact, if $t$ is trajectory time and $s$ arc-length along geodesics, then from \S \ref{s:traj-as-geodesics} and \S \ref{s:jm-metric-config-space-hopf-coords}, $\sigma = ds/dt = \sqrt{2} (E + 3 Gm^3/r^2)$ since $h = 3$. Near a triple collision (small $r$), $ds^2 \approx 3 Gm^3 dr^2/r^2$ so that $s \approx - \half \sqrt{3 Gm^3} \log(1-t/t_c) \to \infty$ as $t \to t_c = r(0)^2/2\sqrt{6 G m^3}$ which is the approximate time to collision. In fact, the exact collision time $t_c = \sqrt{6 G m^3} \left( -1 + \sqrt{1+ {\kappa r(0)^2/6 G m^3}} \right)/\kappa$ may be obtained by reducing Newton's equations for Lagrange homotheties to the one body problem $r^3 \ddot r = - 6 G m^3$ whose conserved energy is $\kappa = \dot r^2 - 6 G m^3 / r^2$. These homothety solutions illustrate how the geodesic flow reformulation regularizes the original Newtonian 3 body dynamics in the inverse-square potential.

More generally, for unequal masses (\ref{e:jm-metric-in-jacobi-coordinates-on-c3})-(\ref{e:Hopf-coords}) give the JM metric $ds^2 = \tl h dr^2/r^2 + \tl g_{i j} dx^i dx^j$ where 
	\beq
	\tl h = \fr{G m_1 m_2 M_1}{\sin^2 \eta} + \fr{G m_2 m_3 M_2}{\lv \cos \eta - \mu_1 \sqrt{{M_2}/{M_1}}e^{2 i \xi_2} \sin \eta \rv^2} + \fr{G m_1 m_3 M_2}{\lv\cos \eta + \mu_2 \sqrt{{M_2}/{M_1}}e^{2 i \xi_2} \sin \eta \rv^2}.
	\label{e:unequal-mass-htilde}
	\eeq
Irrespective of the masses, $\tl g_{ij}$ (\ref{e:kinetic-metric-s3-s2}) is positive and $\tl h$ has a strictly positive lower bound (e.g. $G m_1 m_2 M_1$). Thus by the same argument as above, triple collisions are at infinite distance. Combining this with the corresponding results for pairwise collision points (\S\ref{s:near-pairwise-collision-geometry}), we conclude that the zero-energy JM metrics on $\mC^2$ and $\mR^3$ are geodesically complete for arbitrary masses. 

For non-zero energy, $ds^2 = (E + \tl h/r^2) (dr^2 + r^2 \tl g_{ij} dx^i dx^j)$ which can be approximated with the zero-energy JM metrics both near binary (say, $\eta = 0$) and triple ($r = 0$) collisions. If $\g$ is a curve ending at the triple collision, $l(\g) \geq l(\tl \g)$ where $\tl \g$ is a `tail end' of $\g$ lying in a sufficiently small neighborhood of $r=0$ (i.e., $r \ll |\tl h/E|^{1/2}$ which is guaranteed, say, if $r \ll |G m_1 m_2 M_1/E|^{1/2}$). But then, $l(\tl \g)$ may be estimated using the zero-energy JM metric giving $l(\tl \g) = \infty$. Thus $l(\g) = \infty$. A similar argument shows that curves ending at binary collisions have infinite length. Thus we conclude that the JM metrics on $\mC^2$ and $\mR^3$ are geodesically complete for arbitrary  energies and masses.

\subsection{Scalar curvature for equal masses and zero energy}
\label{s:scalar-curvature-inv-sq-pot-c2-r3-s3-s2}

\begin{figure}	
	\centering
	\begin{subfigure}[t]{2.5in}
		\centering
		\includegraphics[width=3in]{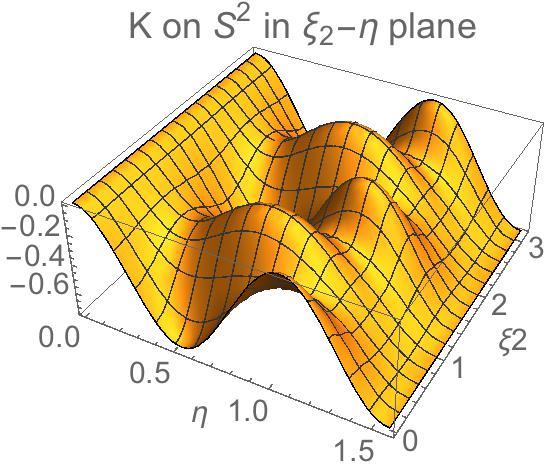}
	\end{subfigure}
	\caption{\small Gaussian curvature $K$ (in units of ${1/Gm^3}$) on $\mS^2$ for equal masses and $E=0$. $K = 0$ at $L_{4,5}$ and $C_{1,2,3}$.}
	\label{f:shape-sphere-curvature}
\end{figure}

\begin{figure}	
	\centering
	\begin{subfigure}[t]{1.9in}
		\centering
		\includegraphics[width=2in]{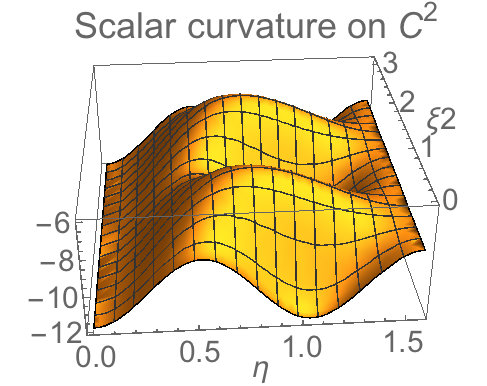}
	\end{subfigure}
	\quad
	\begin{subfigure}[t]{1.9in}
		\centering
		\includegraphics[width=2in]{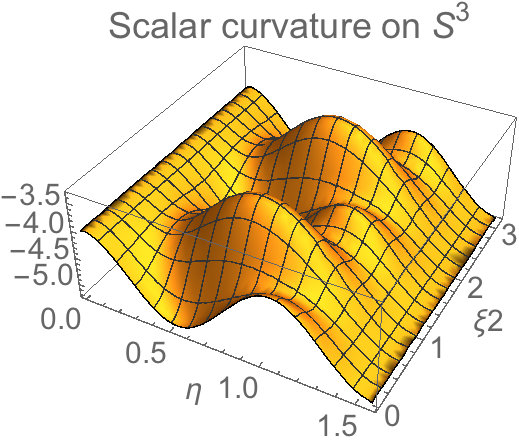}
	\end{subfigure}
	\quad
	\begin{subfigure}[t]{1.9in}
		\centering
		\includegraphics[width=2in]{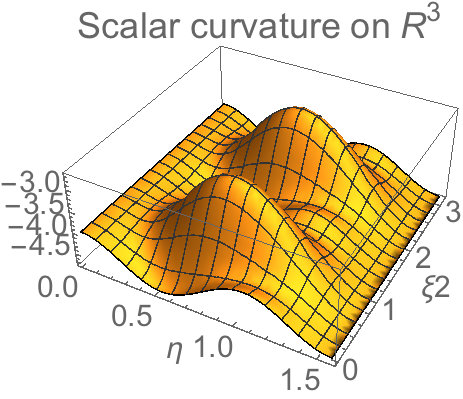}
	\end{subfigure}
\caption{\footnotesize Scalar curvatures $R$ on $\mC^2$, $\mS^3$ and $\mR^3$ in units of $1/{G m^3}$. $R$ is strictly negative and has a global maximum at $L_{4,5}$ in all cases. It attains a global minimum at $C_{1,2,3}$ on $\mC^2$ and a local maximum at collisions on $\mR^3$ and $\mS^3$. $E_{1,2,3}$ are saddles on $\mC^2$ and global minima on $\mR^3$ and $\mS^3$.}
\label{f:scalar-curvature}
\end{figure}

A geodesic through $P$ in the direction $u$ perturbed along $v$ is linearly stable/unstable [see \S \ref{s:stability-tensor}] according as the sectional curvature $K_P(u,v)$ is positive/negative. The scalar curvature $R$ at $P$ is proportional to an average of sectional curvatures in planes through $P$ (\S \ref{s:sectional-curvature-inv-sq-pot}). Thus $R$ encodes an average notion of geodesic stability. Here, we evaluate the scalar curvature $R$ of the equal-mass zero-energy JM metric on $\mC^2$ and its submersions to $\mR^3$, $\mS^3$ and $\mS^2$. In each case, due to the rotation and scaling isometries, $R$ is a function only of the coordinates $\eta$ and $\xi_2$ that parametrize the shape sphere. In \cite{montgomery-pants} Montgomery proves that $R_{\mS^2} \leq 0$ with equality at Lagrange and collision points (see Fig. \ref{f:shape-sphere-curvature}). We generalize this result and prove that the scalar curvatures on $\mC^2$, $\mR^3$ and $\mS^3$ are strictly negative and bounded below (see Fig. \ref{f:scalar-curvature}) indicating widespread linear instability of the geodesic dynamics. (Note that hyperbolicity of the configuration space quotiented by translations, rotations and scaling does not extend in a simple manner to the $4$-body problem \cite{montgomery-jackman}.)

{\fl \bf Scalar curvature on $\mS^2$:} The quotient JM metric on $\mS^2$ (\ref{e:s2-metric-in-eta-xi2-coordinates}) is conformal to the round (kinetic) metric on a sphere of radius $1/2$:
	\beq
	ds_{\mS^2}^2 = Gm^3 \: h(\eta, \xi_2) \: ds_{\rm kin}^2 \quad \text{where} 
	\quad ds_{\rm kin}^2 =
	d\eta^2+ \sin^2 2\eta \: d\xi_2^2.
	\label{e:jm-metric-s2}
	\eeq
Here the conformal factor ($h = - (r^2/Gm^3) \times$ potential energy) (\ref{e:conformal-prefactor-h}) is a strictly positive function on the shape sphere with double poles at collision points. The scalar curvature of (\ref{e:jm-metric-s2}) is
	\beq
	\label{e:scalar-curvature-on-s2}
	R_{\mS^2} =  \ov{ G m^3 h^3 } \left( 8 h^2 + |\grad h|^2 - h \D h \right),
	\eeq	
where $\D$ is the Laplacian and $\grad^i h = g^{ij} \pdr_j h$ the gradient on $\mS^2$ relative to the kinetic metric:
	\beqs
	\label{e:defn-laplace-gradient}
	\D h = \left(\ov{\sin^2 2\eta} \fr{\pdr^2 h}{\pdr \xi_2^2} + 2\cot 2\eta \fr{\pdr h}{\pdr \eta} + \fr{\pdr^2 h}{\pdr \eta^2} \right) \quad \text{and} \quad
	|\grad h|^2 = \fr{1}{\sin^2 2\eta} \left(\fr{\pdr h}{\pdr \xi_2}\right)^2 +  \left(\fr{\pdr h}{\pdr \eta}\right)^2 .
	\label{e:grad-laplace-on-s2-kin}
	\eeqs
In fact we have an explicit formula for the scalar curvature, $R_{\mS^2} = AB/C$ where
	\beqs
	A &=& 8 \sin ^2\eta \left((\cos 2 \eta + 2)^2 - 3 \sin ^2 2 \eta  \cos ^2 2\xi_2 \right), \quad
	C = {3 \left(2 \sin ^2 2 \eta  \cos 4 \xi_2+\cos 4 \eta -13\right)^3} \;\; \& \cr
	B &=& \left(-8 \sin^4 2 \eta \cos 8 \xi_2 - 16 \sin^2 2 \eta   \cos 4 \xi_2 (\cos 4 \eta - 29) + 236 \cos 4 \eta  - 3 \cos 8 \eta + 727 \right).
	\label{e:scalar-curv-shape-sph}
	\eeqs
As shown in \cite{montgomery-pants}, $R_{\mS^2} \leq 0$ with equality only at Lagrange and collision points. Negativity of $R_{\mS^2}$ also follows from (\ref{e:scalar-curv-shape-sph}): each factor in the numerator is $\geq 0$ (the third vanishes at $L_{4,5}$, the second at $C_{1,2}$ and the first at $C_3$) while the denominator is strictly negative. We now use this to show that the scalar curvatures on configuration space $\mC^2$ and its quotients $\mR^3$ and $\mS^3$ are strictly negative.

{\fl \bf Scalar curvature on $\mC^2$:} The equal-mass zero-energy JM metric on $\mC^2$ from Eq. (\ref{e:c2-metric-in-r-eta-xi1-xi2-coordinates}) is
	\beq
	ds^2_{\mC^2}= \left( {G m^3}/{r^2} \right) h(\eta,\xi_2) \left(dr^2+r^2\left(d\eta^2+ d\xi_1^2-2\cos2\eta\;d\xi_1\;d\xi_2+ d\xi_2^2\right)\right).
	\eeq
The scalar curvature of this metric is expressible as
	\beq
	\label{e:scalar-curvature-on-c2}
	R_{\mC^2}= \left( 3/2G m^3h^3 \right) \left(4h^2+ |\grad h|^2 - 
	2 h \: \D h \right),
	\eeq
where $\D h$ and $\grad h$ are the Laplacian and gradient with respect to the {\it round} metric on $\bf S^2$ of radius one-half (\ref{e:grad-laplace-on-s2-kin}). Due to the scaling and rotation isometries, $R_{\mC^2}$ is in fact a function on the shape sphere. The scalar curvatures on $\mC^2$ (\ref{e:scalar-curvature-on-c2}) and $\mS^2$(\ref{e:scalar-curvature-on-s2}) are simply related:
	\beq
	\label{e:scalar-curv-c2-compare-s2}
	R_{\mC^2}=3 R_{\mS^2} -  \left( 3/2 G m^3 h^3 \right) \left( 12 h^2 +|\grad h|^2\right).
	\eeq
This implies $R_{\mC^2} < 0$ since the second term is strictly negative everywhere as we now show. Notice that the second term can vanish only when $h$ is infinite, i.e., at collisions. Taking advantage of the fact that the geometry (on $\mS^2$ and $\mC^2$) in the neighborhood of all 3 collision points is the same for equal masses, it suffices to check that the second term has a strictly negative limit at $C_3$ $(\eta = 0)$. Near $\eta =0$, $h \sim 1/2 \eta^2$ so that $R_{\mC^2} \to - 12/Gm^3 < 0$. Combining with the $r$-independence of $R_{\mC^2}$, we see that the scalar curvature is non-singular at binary and triple collisions.

With a little more effort, we may obtain a non-zero upper bound for the Ricci scalar on $\mC^2$. Indeed, using $R_{\mS^2} \leq 0$ and the inequality $12 h^2 + |\grad h|^2 \geq \zeta h^3$ proved in Appendix \ref{s:upper-bound-scalar-curvature}, we find
	\beq
	R_{\mC^2} < - 3 \zeta/2 G m^3 \quad \text{where} \quad \zeta = 55/27.
	\eeq
Numerically, we estimate the optimal value of $\zeta$ to be $8/3$.

{\bf \fl Scalar curvatures on $\mR^3$ and $\mS^3$:} Recall that the equal-mass zero-energy quotient JM metrics on shape space $\mR^3$ (\ref{e:r3-metric-in-r-eta-xi2-coordinates}) and $\mS^3$ (\ref{e:s3-metric-in-eta-xi1-xi2-coordinates}) are
	\beqs
	ds_{\mR^3}^2 &=&  \left( Gm^3 h/r^2 \right) \left(dr^2+r^2\left(d\eta^2+\sin^22\eta \;d\xi_2^2\right) \right) \quad \text{and} \quad 
	\cr \cr
	ds_{\mS^3}^2 &=& Gm^3 h \: \left(d\eta^2+ d\xi_1^2-2\cos2\eta\;d\xi_1\;d\xi_2+ d\xi_2^2\right). 
	\label{e:r3-s3-jm-metric}
	\eeqs
The corresponding scalar curvatures are
	\beq
	R_{\mR^3}= \left(16h^2+3 |\grad h|^2 - 4 h \D h \right)/2 G m^3h^3
	 \;\; \text{and} \;\; 
	R_{\mS^3}= \left(12h^2+3|\grad h|^2-4h \D h\right)/2 G m^3h^3.
	\eeq
Here $\D h$ and $\grad h$ are as in Eq. (\ref{e:grad-laplace-on-s2-kin}).  The scalar curvatures are related to that on $\mS^2$ as follows
	\beq
	R_{\mR^3}=2 R_{\mS^2} - \left( 16 h^2 +|\grad h|^2\right)/{2 G m^3 h^3} 
	\quad \text{and} \quad R_{\mS^3}=2 R_{\mS^2} - \left( 20 h^2 + |\grad h|^2\right)/{2 G m^3 h^3}.
	\eeq
As in the case of $\mC^2$ we check that the second terms in both relations are strictly negative. This implies both the scalar curvatures are strictly negative. In fact, using the inequality $12 h^2 + |\grad h|^2 > \zeta h^3$ (see Appendix \ref{s:upper-bound-scalar-curvature}) we find (non-optimal) non-zero upper bounds 
	\beq
	R_{\mS^3, \mR^3} < - \zeta/2 G m^3 \quad \text{where} \quad \zeta = 55/27.
	\eeq
Moreover, we note that
	\beq
	R_{\mC^2} = R_{\mS^3} - \frac{h \D h}{G m^3 h^3} < R_{\mS^3} \quad \text{and} \quad
	R_{\mS^3} = R_{\mR^3} - \frac{4 h^2}{2 G m^3 h^3} \leq R_{\mR^3},
	\eeq
with equality at collision configurations. Recalling that on the shape sphere, the scalar curvature vanishes at collision points (in a limiting sense) and at Lagrange points, we have the following inequalities
	\beq
	0 \geq R_{\mS^2} > R_{\mR^3} \geq R_{\mS^3} > R_{\mC^2}.
	\eeq
Thus we have the remarkable result that the scalar curvatures of the JM metric on $\mC^2$ and its quotients by scaling $(\mS^3)$ and rotations $(\mR^3)$ are strictly negative everywhere and also strictly less than that on $\mS^2$. So the full geodesic flow on $\mC^2$ is in a sense more unstable than the corresponding flow on $\mS^2$. 

In addition to strict negativity, we may also show that the scalar curvatures are bounded below. For instance, from Eq. (\ref{e:scalar-curvature-on-s2}) $R_{\mS^2}$ can go to $- \infty$ only when $\D h \to \infty$ since $h \geq 3$. Now from Eq. (\ref{e:defn-laplace-gradient}) $\D h$ can diverge only when $\sin 2 \eta = 0$ or when one of the relevant derivatives of $h$ diverges. From Eq. (\ref{e:conformal-prefactor-h}) this can happen only if $\eta= 0$ (C3) or $\eta = \pi/2$ (E3) or when one of the $v_i \to \infty$, i.e., at collisions. However $\D h = 66$ is finite at $\eta = \pi/2$ and we know from \S \ref{s:near-pairwise-collision-geometry} that $R_{\mS^2}$ is finite at collisions so that $R_{\mS^2}$ is bounded below. The same proof shows that scalar curvatures are bounded below on $\mR^3, \mS^3$ and $\mC^2$ as well.

\subsection{Sectional curvature for three equal masses}
\label{s:sectional-curvature-inv-sq-pot}

In \S \ref{s:scalar-curvature-inv-sq-pot-c2-r3-s3-s2}, we showed that the Ricci scalars $R$ on configuration space and its quotients are negative everywhere, save at Lagrange and collision points on the shape sphere where it vanishes. However, $R$ encodes the stability of geodesics only in an average sense. More precisely, a geodesic through $P$ in the direction $u$ subject to a perturbation along $v$ is linearly stable/unstable according as the sectional curvature $K_P(u,v)$ is positive/negative (see \S \ref{s:stability-tensor}). Here, the sectional curvature which is a function only of the 2-plane spanned by $u$ and $v$ generalizes the Gaussian curvature to higher dimensions. It is defined as the ratio of the curvature biquadratic $\scripty{r} = g(R(u,v)v,u)$ to the square of the area ${\rm Ar}(u,v)^2 =  g(u,u) g(v,v) - g(u,v) g(v,u)$ of the parallelogram spanned by $u$ and $v$. Here $g(u,v)$ is the Riemannian inner product and $R(u,v) = [\grad_u, \grad_v] - \grad_{[u,v]}$ the curvature tensor with components $R(e_i, e_j) e_k = R^l_{\; k ij} e_l$ in any basis for vector fields. Furthermore, if $e_1, \ldots, e_n$ are an orthonormal basis for the tangent space at $P$, then  the scalar curvature $R = \sum_{i \ne j} K(e_i,e_j)$ is the sum of sectional curvatures in $n \choose 2$ planes through $P$. It may also be regarded as an average of the curvature biquadratic $R = \iint \scripty{r}(u,v) d\mu_g(u) d\mu_g(v)$ where $d\mu_g(u) = \exp \left(- u^i u^j g_{ij}/2 \right) du$ is the gaussian measure on tangent vectors with mean zero and covariance $g^{ij}$ \cite{Rajeev-geometry-fluid-rigid}. Thus $R$ provides an averaged notion of stability. To get a more precise measure of linear stability of geodesics we find the sectional curvatures in various (coordinate) tangent $2$-planes of the configuration space and its quotients. On account of the isometries, these sectional curvatures are functions only of $\eta$ and $\xi_2$ [explicit expressions are omitted due to their length]. Unlike scalar curvatures which were shown to be non-positive, we find planes in which sectional curvatures are non-positive as well as planes where they can have either sign.

O'Neill's theorem allows us to determine or bound certain sectional curvatures on the configuration space $\mC^2$ in terms of the more easily determined curvatures on its quotients. Roughly, the sectional curvature of a horizontal two-plane increases under a Riemannian submersion. Suppose $f: (M,g) \to (N,\tilde g)$ is a Riemannian submersion. Then O'Neill's theorem \cite{oneill} states that the sectional curvature in any horizontal $2$-plane at $m \in M$ is less than or equal to that on the corresponding $2$-plane at $f(m) \in N$:
	\beq
	\label{e:oneill}
	K_N(df(X),df(Y))=K_M(X,Y)+\frac{3}{4}\fr{|[X,Y]^V|^2}{{\rm Ar}(X,Y)^2}.
	\eeq
Here $X$ and $Y$ are horizontal fields on $M$ spanning a non-degenerate $2$-plane (${\rm Ar}(X,Y)^2 \ne 0$) and $[X,Y]^V$ is the vertical projection of their Lie bracket. In particular, the sectional curvatures are equal everywhere if $X$ and $Y$ are coordinate vector fields.

\begin{figure}	
	\centering
	\begin{subfigure}[t]{1.9in}
		\centering
		\includegraphics[width=2in]{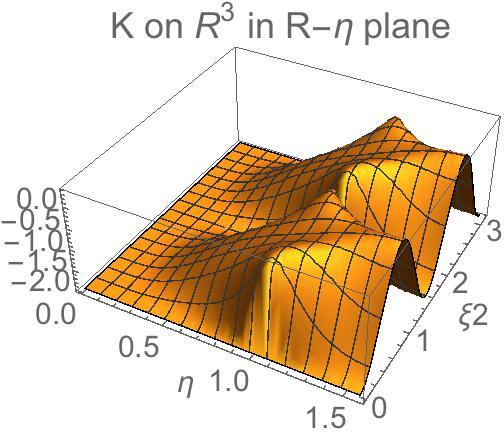}
		\caption{$K_{\mC^2}(\pdr_r,\pdr_\eta) = K_{\mR^3}(\pdr_r,\pdr_\eta) \leq 0$ everywhere except in neighborhoods of $E_3$. $K = -2$ at its global minimum $C_3$ and $K=-2/3$ at $L_{4,5}$. $K \to 0,-2$ when $C_{1,2}$ are approached holding $\eta$ or $\xi_2$ fixed.}
	\end{subfigure}
	\quad
	\begin{subfigure}[t]{1.9in}
		\centering
		\includegraphics[width=2in]{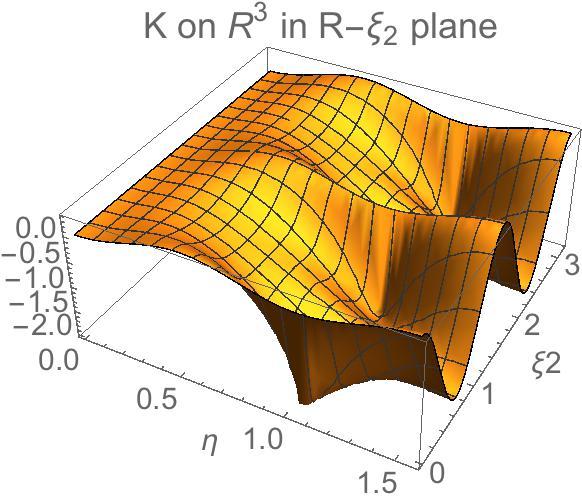}
		\caption{$K_{\mC^2}(\pdr_r, \pdr_\xi) = K_{\mR^3}(\pdr_r, \pdr_{\xi_2})$ is negative except in neighborhoods of $C_3$ and $E_3$. $K = 0$ at its minimum $C_3$ $(\eta = 0)$ and $K=-2/3$ at $L_{4,5}$. $K \to -2$ or $0$ on approaching $C_{1,2}$ $(\eta = \pi/3, \xi_2 = 0, \pi/2)$ along $\eta$ or $\xi_2$ constant.}
	\end{subfigure}
		\quad
	\begin{subfigure}[t]{1.9in}
		\centering
		\includegraphics[width=2in]{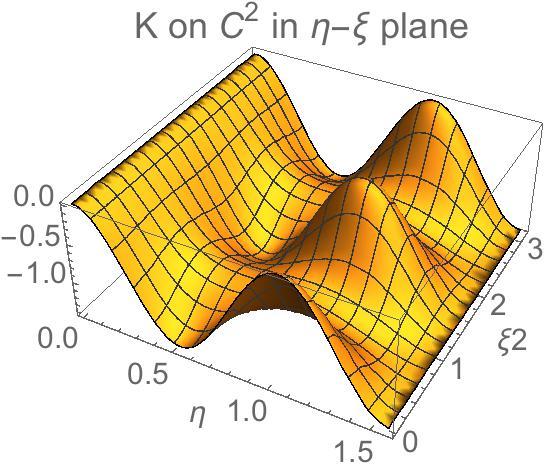}
		\caption{$K_{\mC^2}(\pdr_\eta, \pdr_{\xi}) \leq K_{\mR^3}(\pdr_\eta, \pdr_{\xi_2})$. $K_{\mC^2}(\pdr_\eta, \pdr_{\xi}) = 0$ at global maxima $C_{1,2,3}$ and is negative elsewhere. $K=-1$ at its local maxima $L_{4,5}$.}
	\end{subfigure}
\caption{Sectional curvatures on horizontal 2-planes of submersion from $\mC^2$ to $\mR^3$ in units of $1/{Gm^3}$.}
\label{f:shape-space-curvature}
\end{figure}

We consider sectional curvatures in $6$ interesting $2$ planes on $\mC^2$ which are horizontal with respect to submersions to $\mR^3$ and $\mS^3$. Under the submersion from $\mC^2$ to $\mR^3$ (\S \ref{s:quotient-metrics}), the horizontal basis vectors $\pdr_r$, $\pdr_\eta$ and $\pdr_\xi \equiv \cos 2 \eta \pdr_{\xi_1} + \pdr_{\xi_2}$ map respectively to $\pdr_r$, $\pdr_\eta$ and $\pdr_{\xi_2}$ defining three pairs of corresponding $2$-planes. Since $[\pdr_r, \pdr_\eta] $ and $[\pdr_r,\pdr_\xi]$ vanish, we have $K_{\mC^2}(\pdr_r,\pdr_\eta) = K_{\mR^3}(\pdr_r,\pdr_\eta)$ and  $K_{\mC^2}(\pdr_r,\pdr_\xi) = K_{\mR^3}(\pdr_r,\pdr_{\xi_2})$. Fig. \ref{f:shape-space-curvature} shows that $K_{\mC^2}(\pdr_r,\pdr_\eta)$ is mostly negative, though it is not continuous at $E_3$, $C_1$ and $C_2$. On the other hand $K_{\mC^2}(\pdr_r,\pdr_{\xi})$ is largely negative except in a neighbourhood of $C_3$. Finally, as $[\pdr_\xi,\pdr_\eta]^V = - 2 \sin 2 \eta \pdr_{\xi_1} \ne 0$, we have $K_{\mC^2}(\pdr_\eta,\pdr_\xi) < K_{\mR^3}(\pdr_\eta,\pdr_{\xi_2})$ with equality at collisions. Moreover the submersion from $\mR^3 \to \mS^2$ (\S \ref{s:quotient-metrics}) implies that $K_{\mR^3}(\pdr_\eta,\pdr_{\xi_2})$ coincides with $K_{\mS^2}(\pdr_\eta,\pdr_{\xi_2})$ which vanishes at Lagrange and collision points and is strictly negative elsewhere (see \S \ref{s:scalar-curvature-inv-sq-pot-c2-r3-s3-s2}). Thus $K_{\mC^2}(\pdr_\eta,\pdr_\xi)$ vanishes at collision points and is strictly negative everywhere else (see Fig. \ref{f:shape-space-curvature}). In particular, Lagrange points are more unstable on the configuration space $\mC^2$ than on the shape sphere.

\begin{figure}	
	\centering
	\begin{subfigure}[t]{1.9in}
		\centering
		\includegraphics[width=2in]{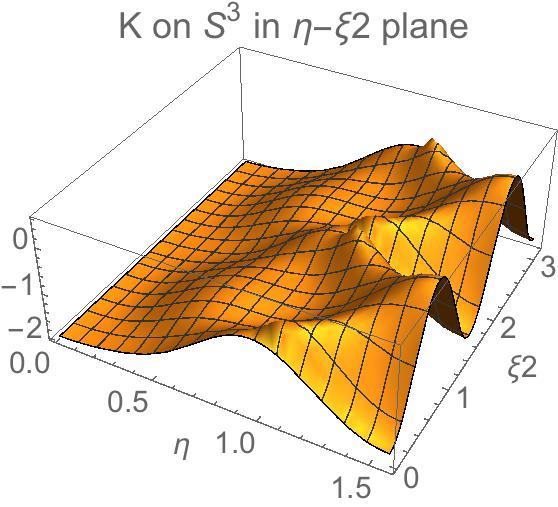}
		\caption{$K_{\mC^2}(\pdr_\eta, \pdr_{\xi_2}) = K_{\mS^3}(\pdr_\eta, \pdr_{\xi_2}) > 0$ in a neighbourhood of $E_3$ and negative elsewhere. $K = -2$ at its global minimum $C_3$. $K = -1$ at its local maxima $L_{4,5}$. $K \to$ $0$ or $-1/2$ upon approaching $C_{1,2}$ along constant $\eta$ or $\xi_2$.}
	\end{subfigure}
	\quad
	\begin{subfigure}[t]{1.9in}
		\centering
		\includegraphics[width=2in]{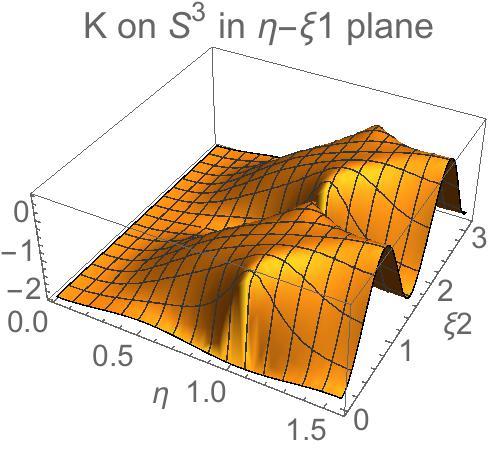}
		\caption{$K_{\mC^2}(\pdr_\eta, \pdr_{\xi_1}) = K_{\mS^3}(\pdr_\eta, \pdr_{\xi_1}) > 0$ in a neighbourhood of $E_3$ and is negative elsewhere. $K= -2$ at its global minimum $C_3$ and $K = -1/3$ at $L_{4,5}$. $K \to 0$ or $-2$ upon approaching $C_{1,2}$ holding $\eta$ or $\xi_2$ fixed.}
	\end{subfigure}
	\quad
	\begin{subfigure}[t]{1.9in}
		\centering
		\includegraphics[width=2in]{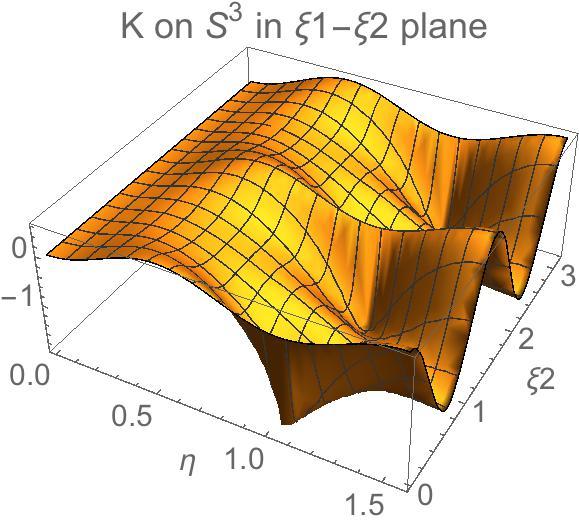}
		\caption{$K_{\mC^2}(\pdr_{\xi_1}, \pdr_{\xi_2}) = K_{\mS^3}(\pdr_{\xi_1}, \pdr_{\xi_2}) > 0$ in some neighbourhoods of $C_3$ and $E_{3}$ and negative elsewhere. $K = 0$ at its local minimum $C_3$. $K=-1/3$ at $L_{4,5}$. $K \to -2$ or $0$ upon approaching $C_{1,2}$ while holding $\eta$ or $\xi_2$ fixed.}
	\end{subfigure}
\caption{\small Sectional curvatures on horizontal 2-planes of submersion from $\mC^2$ to $\mS^3$ in units of $1/{Gm^3}$.}
\label{f:s3-curvature}
\end{figure}

Under the submersion from $\mC^2$ to $\mS^3$ (\S \ref{s:quotient-metrics}), the horizontal basis vectors $\pdr_\eta$, $\pdr_{\xi_1}$ and $\pdr_{\xi_2}$ map respectively to $\pdr_\eta$, $\pdr_{\xi_1}$ and $\pdr_{\xi_2}$. The sectional curvatures on corresponding pairs of 2-planes are equal, e.g. $K_{\mC^2}(\pdr_\eta,\pdr_{\xi_2}) =K_{\mS^3}(\pdr_\eta,\pdr_{\xi_2})$. 
As shown in Fig. \ref{f:s3-curvature}, $K_{\mC^2}(\pdr_\eta,\pdr_{\xi_2})$ is negative everywhere except in a neighbourhood of $E_3$ where it can have either sign. The qualitative behavior of the other two sectional curvatures $K_{\mC^2}(\pdr_{\xi_1},\pdr_{\xi_2})$ and $K_{\mC^2}(\pdr_{\xi_1}, \pdr_\eta)$  is similar to that of $K_{\mC^2}(\pdr_{r},\pdr_{\xi_2})$ and $K_{\mC^2}(\pdr_{r}, \pdr_\eta)$ discussed above. The approximate symmetry under $\pdr_{\xi_1} \leftrightarrow \pdr_r$ is not entirely surprising given that $\pdr_{\xi_1}$ and $\pdr_{r}$ are vertical vectors in the submersions to $\mR^3$ and $\mS^3$ respectively.
   
The remaining two coordinate 2-planes on $\mC^2$ are not horizontal under either submersion. We find that $K_{\mC^2}(\pdr_r, \pdr_{\xi_1})$ is negative everywhere except at $L_{4,5}$ and $K_{\mC^2}(\pdr_r, \pdr_{\xi_2})$ is negative except around $E_{1,2}$.

\subsection{Stability tensor and linear stability of geodesics}
\label{s:stability-tensor}

In this section we use the stability tensor (which provides a criterion for linear geodesic stability) to discuss the stability of Lagrange rotational and homothety solutions. We end with a remark on linear stability of trajectories and geodesics. Consider the $n$-dimensional configuration manifold $M$ with metric $g$. The geodesic deviation equation (GDE) for the evolution of the separating vector (Jacobi field) $y(t)$ between a geodesic $x(t)$ and a neighboring geodesic is \cite{oneill}
	\beq
	\label{e:geodesic-deviation-eqn-jacobi-field}
	\grad_{\dot x}^2 y = R(\dot x, y) \dot x = - R(y, \dot x) \dot x.
	\eeq
We expand the Jacobi field $y=c^k(t) e_k(t)$ in any basis $e_i(t)$ that is parallel transported along the geodesic i.e. $\grad_{\dot x}e_k=0$ [$e_i(0)$ could be taken as coordinate vector fields at $x(0)$]. Taking the inner product of the GDE with $e_m$ and contracting with $g^{im}$, we get $\ddot c^i = - S^i_j c^j$, where the `stability tensor' $S^i_k = R^i_{ jkl}\dot x^j \dot x^l$. As $S$ is real symmetric, its eigenvectors $f_i$ can be chosen to form an orthonormal basis for $T_x M$. Writing $y = d^m f_m$, the GDE becomes $\ddot d^m = -\kappa_m d^m$ (no sum on $m$) where $\kappa_m$ is the eigenvalue of $S$ corresponding to the eigenvector $f_m$. The eigenvalues of $S$ (say at $t = 0$) control the initial evolution of the Jacobi fields in the corresponding eigendirections. Since $\kappa_m = \left( {\rm Area}{\bra f_k , \dot x \ket} \right)^2 K(f_m, \dot x)$ (\S \ref{s:sectional-curvature-inv-sq-pot}), positive (negative) $\kappa$ or $K$ imply local stability (instability) for the initial evolution. We note that calculating $S$ and its eigenvalues at a given instant (say $t=0$) requires no knowledge of the time evolution of $e_i(t)$. So we may simply use the coordinate vector fields as the basis. Notice that the tangent vector to the geodesic $\dot x$ is always an eigendirection of $S$ with eigenvalue zero.

{\fl \bf Rotational Lagrange solutions in Newtonian potential:} Consider the Lagrange rotational solutions where three equal masses ($m_i = m$) rotate at angular speed $\om = \sqrt{3 G m / a^3}$ around their CM at the vertices of an equilateral triangle of side $a$. The rotational trajectory on $\mC^2$ in $r,\eta,\xi_{1,2}$ coordinates is given by $x(t) = (a/\sqrt{m}, \pi/4, \om t, \pm \pi/4)$ with velocity vector $\om \pdr_{\xi_1}$. Note that trajectory and geodesic times are proportional since $\sigma = ds/dt = (E-V)/\sqrt{\cal T}$ with $V(r, \eta, \xi_2)$ and $\cal T$ constant along $x(t)$. The stability tensor along the geodesic, $S = \om^2 \; \text{diag} (1,-1/2, 0, -1/2 )$ is diagonal in the coordinate basis $r,\eta, \xi_1, \xi_2$. As always, $\dot x$ is a zero-mode. A perturbation along $\pdr_r$ is linearly stable while those directed along $\pdr_\eta$ or $\pdr_{\xi_2}$ are linearly unstable. Note that Routh's criterion $27 (m_1 m_2 + m_2 m_3 + m_3 m_1) < M^2$ \cite{routh} predicts that Lagrange rotational solutions are linearly unstable for equal masses.

{\fl \bf Lagrange homotheties:} For equal masses, a Lagrange homothety solution is one where the masses move radially (towards/away from their CM) while being at the vertices of equilateral triangles. The geodesic in Hopf coordinates takes the form $(r(t), \eta = \pi/4, \xi_1, \xi_2 = \pm \pi/4)$ where $\xi_1$ is arbitrary and independent of time. Though an explicit expression is not needed here, $r(t)$ is the solution of $\ddot r + \G^r_{rr} \dot r^2 = 0$ where $\G^r_{rr} = - 3 G m^3/(E r^3 + 3 Gm^3 r)$ for the inverse-square potential. The stability tensor is diagonal:
	\beq
	S = \fr{6 G m^3 \dot r^2}{\left( 3 G m^3 r + E r^3 \right)^2}\text{diag} \left(0,- 3 G m^3 - 2 E r^2, - E r^2 , - 3 G m^3 - 2 E r^2 \right).
	\eeq
For a given $r$ and positive energy, perturbations along $\pdr_{\xi_{1,2}}$ and $\pdr_\eta$ are unstable while they are stable when $-3Gm^3/r^2 < E < - 3 Gm^3/2r^2$. For intermediate (negative) energies, $\pdr_{\eta}$ and $\pdr_{\xi_2}$ are unstable directions while $\pdr_{\xi_1}$ is stable. For the Newtonian potential, we have similar conclusions following from the corresponding stability tensor:
	\beq
	S = \fr{3 G m^{5/2} \dot r^2}{4 r^2 \left( 3 G m^{5/2} + E r \right)^2} \text{diag} \left(0,- 9 G m^{5/2} - 5 E  r, - 2 E r, - 9 G m^{5/2} - 5 E  r \right).
	\eeq
We end this section with a cautionary remark. For a system whose trajectories can be regarded as geodesics of the JM metric, linear stability of geodesics may not coincide with linear stability of corresponding trajectories. This may be due to the reparametrization of time (see \S \ref{s:triple-collision-inv-sq-near-collision-geom} for examples) as well as the restriction to energy conserving perturbations in the GDE. We illustrate this with a 2D isotropic oscillator with spring constant $k$. Here the curvature of the JM metric (see \S \ref{s:traj-as-geodesics}) is $R = 2Ek/T^3$ where $T$ is the kinetic energy. Thus for positive $k$, geodesics are always linearly stable while for negative $k$ they are stable/unstable according as energy is negative/positive. By contrast, linearizing the EOM $\ddot \del x_i = - (k/m) \del x_i$ shows that trajectories are linearly stable for positive $k$ and linearly unstable for negative $k$. This (possibly atypical) example illustrates the fact that geodesic stability does not necessarily imply stability of trajectories. 

\section{Planar three-body problem with Newtonian potential}
\label{s:three-body-coulomb}

\subsection{JM metric and its curvature on configuration and shape space}
\label{s:curvature-newtonian-potential}

In analogy with our geometric treatment of the planar motion of three masses subject to inverse-square potentials, we briefly discuss the gravitational analogue with Newtonian potentials. As before, the translation invariance of the Lagrangian 
	\beq
	L = \half \sum_{i=1,2,3} m_i \dot x_i^2 - \sum_{i < j} \fr{G m_i m_j}{|x_i - x_j|}
	\eeq
allows us to go from the configuration space $\mC^3$ to its quotient $\mC^2$ endowed with the JM metric
	\beq
	ds^2 = \left( E + \fr{G m_1 m_2}{|J_1|} + \fr{G m_2 m_3}{|J_2 - \mu_1 J_1|} + \fr{G m_3 m_1}{|J_2+\mu_2 J_1|} \right) \left( M_1 |dJ_1|^2 + M_2 |dJ_2|^2  \right).
	\eeq
The Jacobi coordinates $J_{1,2}$, mass ratios $\mu_{1,2}$ and reduced masses $M_{1,2}$ are as defined in Eqs. (\ref{e:jacobi-coordinates-on-c3}, \ref{e:ke-in-jacobi-coordinates-on-c3}, \ref{e:jm-metric-in-jacobi-coordinates-on-c3}). In rescaled Jacobi coordinates  $z_i = \sqrt{M_i} \:  J_i$ (\ref{e:jm-metric-in-jacobi-coordinates-on-c2}), the JM metric on $\mC^2$ for {\it equal masses} becomes
	\beq
	ds^2 = \left(E+\fr{G m^{5/2}}{\sqrt{2}|z_1|}+\fr{\sqrt{2} G m^{5/2}}{\sqrt{3}|z_2-\ov{\sqrt{3}}z_1|}+\fr{\sqrt{2} G m^{5/2}}{\sqrt{3}|z_2+\ov{\sqrt{3}}z_1|}\right) \left(|dz_1|^2+|dz_2|^2 \right).
	\eeq
Rotations $z_j \mapsto e^{i \tht} z_j$ continue to act as isometries  corresponding to the KVF $\pdr_{\xi_1}$ in Hopf coordinates (\ref{e:Hopf-coords}), where the JM metric is
	\beqs
	\label{e:c2-jm-metric-coulomb}
	ds^2&=& \left(E + {Gm^{5/2}U}/{r}\right) \left(dr^2+r^2\left(d\eta^2+ d\xi_1^2-2\cos2\eta\;d\xi_1\;d\xi_2+ d\xi_2^2\right)\right) \quad \text{with}\cr
	 U&=&\fr{1}{\sqrt{2}\sin\eta} + \fr{\sqrt{2}}{\sqrt{2 + \cos 2 \eta - \sqrt{3} \sin 2 \eta \cos 2 \xi_2}} + \fr{\sqrt{2}}{\sqrt{2 + \cos 2 \eta + \sqrt{3} \sin 2 \eta \cos  2 \xi_2 }}.
	\eeqs
Requiring the submersion $(r,\eta,\xi_1, \xi_2) \mapsto (r,\eta,\xi_2)$ from $\mC^2$ to its quotient by rotations to be Riemannian gives us the JM metric on shape space $\mR^3$:
	\beq
	\label{e:r3-jm-metric-coulomb}
ds^2=  \left(E + {Gm^{5/2}U}/{r}\right) \left(dr^2+r^2\left(d\eta^2+\sin^2 2\eta \;d\xi_2^2\right)\right). 
	\eeq
Unlike for the inverse-square potential, scaling $r \mapsto \la r$ is not an isometry of the JM metric even when $E = 0$. Thus we do not have a further submersion to the shape sphere. However, in what follows, we will consider $E=0$, as it leads to substantially simpler curvature formulae.
 	
\begin{figure}	
	\centering
	\begin{subfigure}[t]{3in}
		\centering
		\includegraphics[width=2.9in]{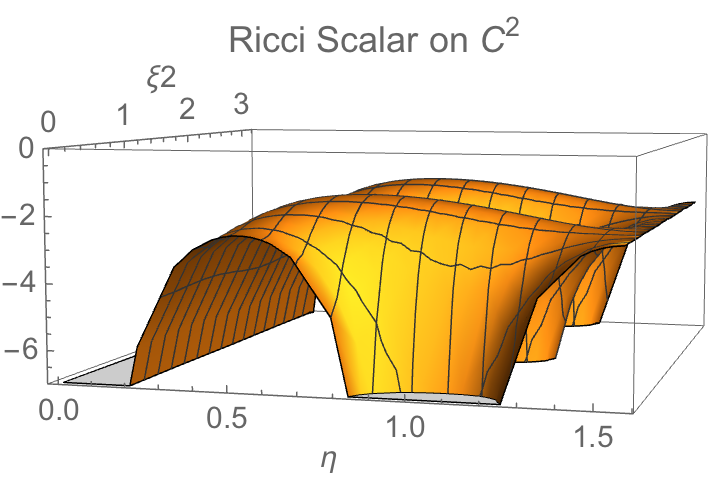}
		\caption{}
	\end{subfigure}
	\quad
	\begin{subfigure}[t]{3in}
		\centering
		\includegraphics[width=3.5in]{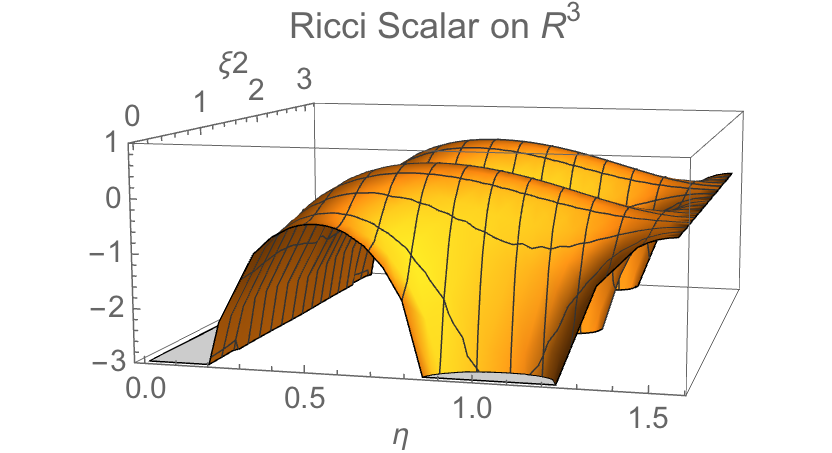}
		\caption{}
	\end{subfigure}
\caption{\footnotesize Ricci scalar $R$ for zero energy and equal masses on $\mC^2$ and $\mR^3$ for the Newtonian potential (in units of ${1/G m^{5/2} r}$). $R$ on $\mC^2$ is strictly negative while that on $\mR^3$ can have either sign.}
\label{f:scalar-curvature-inv-r}
\end{figure}

Though we do not have a submersion to the shape sphere, the quantity $U(\eta, \xi_2)$ in the conformal factor may be regarded as a function on a $2$-sphere of radius one-half. This allows us to express the scalar curvatures as 
	\beq
	\label{e:scalar-curvatures-coulomb}
	R_{\mC^2} = \frac{3}{2 G m^{5/2} r U ^3}
 \left(3 U ^2+ |\grad U|^2-2 U \D U \right) \;\; \text{and} \;\; 
	R_{\mR^3} = \frac{1}{4 Gm^{5/2} r U^3} \left(30 U^2+6|\grad U|^2 -8  U  \D U  \right)
	\eeq
where $\D U$ is the Laplacian and $\grad U$ the gradient relative to the round metric on a $2$-sphere of radius $1/2$. Evidently, both the scalar curvatures vanish in the limit $r \to \infty$ of large moment of inertia $I_{\rm CM} = r^2$; they are plotted in Fig. \ref{f:scalar-curvature-inv-r}. Numerically, we find that for any fixed $r$, $R_{\mC^2}$ is strictly negative and reaches its global maximum $-3/(2 Gm^{5/2} r)$ at the Lagrange configurations $L_{4,5}$, while $R_{\mR^3}$ has a positive global maximum $1/(2 Gm^{5/2} r)$ at the same locations. Note that $R_{\mR^3} = 2 R_{\mC^2}/3 + (9U^2 + |\grad U|^2 )/(2 Gm^{5/2} r U^3)$. As argued in Eq. (\ref{e:scalar-curv-c2-compare-s2}), the second term is strictly positive and vanishes only when $r \to \infty$. Using the negativity of $R_{\mC^2}$, it follows that $R_{\mR^3} > R_{\mC^2}$ with $(R_{\mR^3} - R_{\mC^2})$ attaining its minimum $2/(Gm^{5/2} r)$ at $L_{4,5}$. Thus in a sense, the geodesic dynamics on $\mC^2$ is more linearly unstable than on shape space.  Like the Ricci scalars, sectional curvatures on coordinate $2$-planes are $(1/r) \times $ a function of $\eta$ and $\xi_2$. We find that sectional curvatures are largely negative and often go to $\pm \infty$ at collision points (see Eq. (\ref{e:sec-curv-near-collision-coulomb-pot})).

\subsection{Near-collision geometry and `geodesic incompleteness'}
\label{s:geodesic-incompleteness-newtonian-pot}

Unlike for the inverse-square potential, the scalar curvatures on $\mC^2$ and $\mR^3$ (\ref{e:scalar-curvatures-coulomb}) diverge at binary and triple collisions. To examine the geometry near pairwise collisions of equal masses, it suffices to study the geometry near $C_3$ ($\eta = 0$, $r \ne 0$, $\xi_{1,2}$ arbitrary) which represents a collision of $m_1$ and $m_2$. We do so by retaining only those terms in the expansion of the zero-energy metrics around $\eta=0$:
	\beqs
	\label{e:near-collision-1/r-c2}
	ds_{\mC^2}^2 &\approx& \left( {G m^{5/2}}/{\sqrt{2}\eta r} \right) \left( dr^2+ r^2  \left(d\eta^2+d\xi_1^2-2(1-2\eta^2) d\xi_1 d\xi_2+d\xi_2^2\right)\right) \quad \text{and} \cr
	ds_{\mR^3}^2 &\approx& \left( {G m^{5/2}}/{r}  \right) \left( 1/{\sqrt{2} \eta}+2\sqrt{{2}/{3}}\right) \left( dr^2+ r^2  \left(d\eta^2+4\eta^2d\xi_2^2\right)\right),
	\eeqs
that are necessary to arrive at the following curvatures to leading order in $\eta$:
	\beqs
	\label{e:sec-curv-near-collision-coulomb-pot}
	\text{on $\mC^2$:} && R= {-3}/{\varrho} \;\; \text{and} \;\; 
	K(\pdr_\eta,\pdr_{r,\xi_{1,2}}) = 2 K(\pdr_r,\pdr_{\xi_{1,2}}) = - 2K(\pdr_{\xi_1},\pdr_{\xi_2}) =-{1}/{\varrho } \cr
	\text{on $\mR^3$:} && R = {-1}/{\varrho} \;\; \text{and} \;
\;
	K(\pdr_\eta,\pdr_r) = -2 K(\pdr_r,\pdr_{\xi_2}) = -{1}/{\varrho }, \quad 
	 K(\pdr_\eta,\pdr_{\xi_2}) = -\fr{2\sqrt{{2}/{3}}}{Gm^{5/2}}
	\eeqs
where $\varrho = \sqrt{2} G m^{5/2}\eta r$. The curvature singularity at $\eta = 0$ is evident in the simple poles in the Ricci scalars and all but one of the sectional curvatures in coordinate planes.

We use the near-collision JM metric of Eq. (\ref{e:near-collision-1/r-c2}) to show that a pairwise collision point lies at finite geodesic distance from another point in its neighborhood. Thus, unlike for the inverse-square potential, the geodesic reformulation {\it does not} regularize the gravitational three-body problem. Consider a point $P$ near $\eta = 0$ with coordinates $(r,\eta_0,\xi_1, \xi_2)$. We estimate its distance to the collision point $C_3$ $(r,0,\xi_1, \xi_2)$. To do so, we consider a curve $\g$ of constant $r$, $\xi_1$ and $\xi_2$ running from $P$ to $C_3$ parametrized by $\eta_0 \geq \eta \geq 0$. We will show that $\g$ has finite length so that the geodesic distance to $C_3$ must be finite. In fact, from (\ref{e:near-collision-1/r-c2}):
	\beq
	\text{Length}(\g) = \int_{\eta_0}^{0} \sqrt{\fr{G r m^{5/2}}{\sqrt{2}}}  \fr{d\eta}{\sqrt{\eta}} = - 2 \sqrt{\fr{G r m^{5/2}}{\sqrt{2}}} \sqrt{\eta_0} < \infty.
	\eeq
Furthermore, the image of $\g$ under the Riemannian submersion to shape space $\mR^3$ is a curve of even shorter length ending at a collision point. Thus geodesics on $\mC^2$ and $\mR^3$ can reach binary collisions in finite time, where the scalar curvature is singular. It is therefore interesting to study regularizations of collisions in the three body problem and their geometric interpretation.

{\fl \bf Acknowledgements:} We thank K G Arun, A Lakshminarayan, R Montgomery, S G Rajeev and A Thyagaraja for useful discussions and references. This work was supported in part by the Infosys Foundation and a Ramanujan grant of the Department of Science \& Technology, Govt. of India.

\appendix

\section{Proof of an inequality to give an upper bound for the scalar curvature}
\label{s:upper-bound-scalar-curvature}
 
Here we establish a strict lower bound on the quantity that appears in the relation (\ref{e:scalar-curv-c2-compare-s2}) between Ricci scalars on $\mC^2$ and $\mS^2$. Since Montgomery has shown that $R_{\mS^2} \leq 0$, this helps us  establish strictly negative upper bounds for the scalar curvatures on $\mC^2$, $\mR^3$ and $\mS^3$. We will show here that
	\beq
	\label{e:inequality-zeta}
	 12 h^2 + |\grad h|^2  > \zeta h^3 \quad \text{where} \quad  \zeta = {55}/{27} \approx 2.04.
	\eeq 
The best possible $\zeta$ is estimated numerically to be $\zeta = 8/3$ and the minimum occurs at the Euler points $E_{1,2,3}$. We define the power sum symmetric functions $u_{2n} = \sum_{i=1}^3 v_i^n$ in terms of which the pre-factor in the JM metric (\ref{e:conformal-prefactor-h}) is $h = v_1 + v_2 + v_3 = u_2$. In \cite{montgomery-pants} Montgomery shows that $|\grad h|^2 = 4s$ where the symmetric polynomial
	\beq
	s = (1/2) \left(-2 u_2^2 + 4 u_2 u_4 - 3u_4^2 + 3 u_8 \right).
	\eeq
This gives 
	\beq
	\label{e:A-B}
	12 h^2 + |\grad h|^2 = u_2^3 \left( 8 A + 6 B\right) \quad \text{where} \quad 	A = \fr{u_2 + u_4}{ u_2^2} \quad \text{and} \quad B = \fr{u_8- u_4^2}{u_2^3} .
	\eeq
We will show below that	$ A \geq 17/27$ and $B > -1/2$, from which Eq. (\ref{e:inequality-zeta}) follows (numerically we find that $B \geq -32/81$ which leads to the above-mentioned optimal value $\zeta = 8/3$). To prove the inequality for $B$, we define $c = \cos 2\eta$ and $s = \sin 2\eta \cos 2\xi_2$ which lie in the interval $[-1,1]$. Then  
	\beq
	\fr{u_8 - u_4^2}{u_2^3} > -\half \quad \Leftrightarrow \quad
u_8 - u_4^2 + \fr{u_2^3}{2} > 0 \quad	\Leftrightarrow \quad
	 \frac{3}{8} \left(20 - 3 (c^2 + s^2)^2 - 8 c^3 + 24 c s^2 \right) > 0.
	\eeq
For the latter to hold it is sufficient that $17 - 8 c^3 + 24 c s^2 > 0$ which is clearly true for $0 \leq c \le 1$. For $-1 \leq c < 0$ put $c = -d$. Then it is enough to show that $17 + 8 d^3 - 24 d (1-d^2) > 0$ since $s^2 \leq 1 - d^2$. This holds as the LHS is positive at its boundary points $d = 0, 1$ as well as at its local extremum $d = 1/2$.

\begin{figure}[h]
	\centering
	\includegraphics[width=7cm]{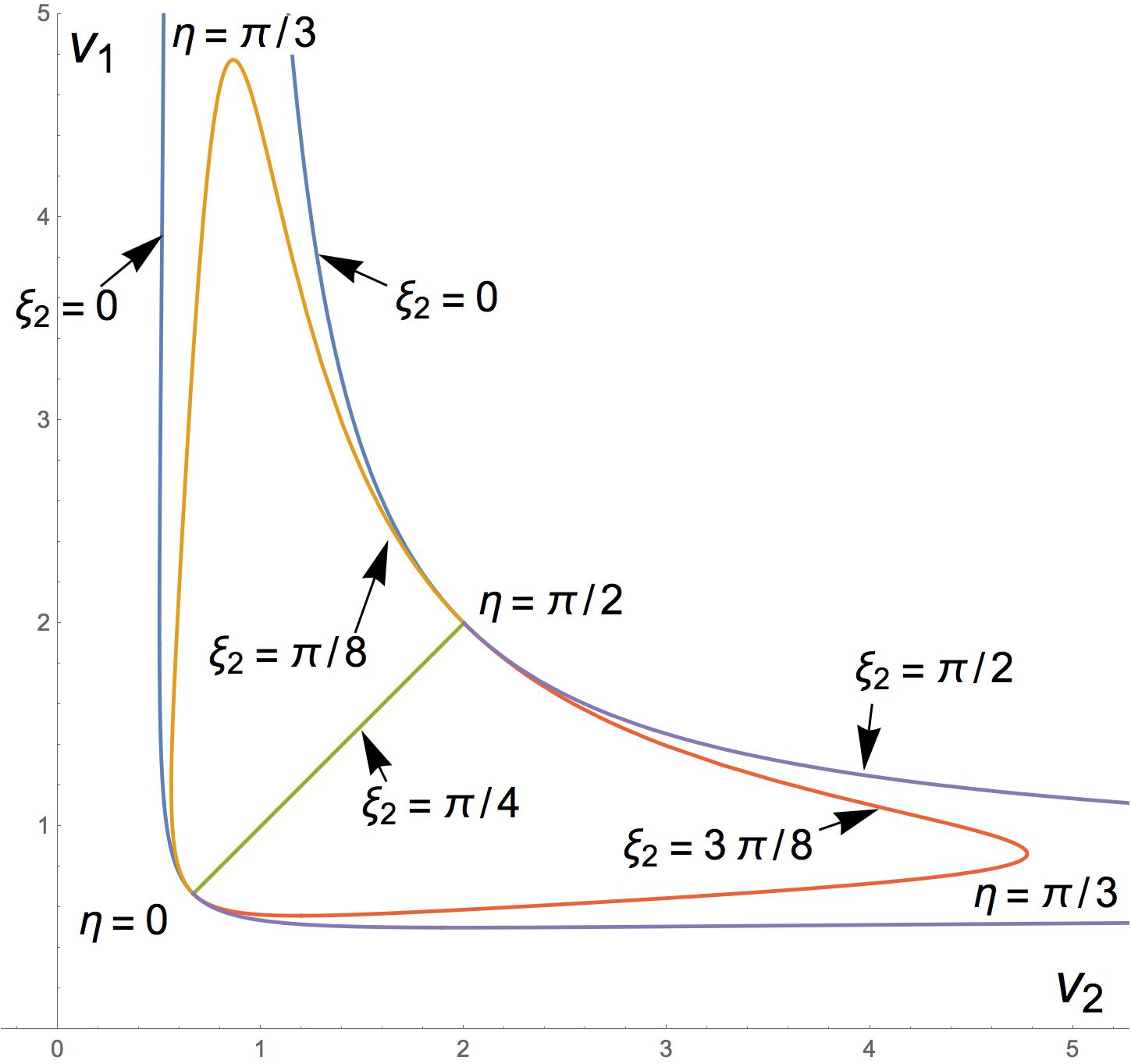}
	\caption{\footnotesize The boundary $\pdr D$ of the region $D$ in the $v_1$-$v_2$ plane is given by the level curves $\xi_2 = 0, \pi/2$. These level curves run from the collision point $\eta = 0$ to the Euler point $\eta = \pi/2$, passing through the collision points at $v_1 = \infty$ or $v_2 = \infty$ (where $\eta = \pi/3$). The level curves $\xi_2 = \pi/8, \pi/4, 3\pi/8$ in the interior $D$ are also shown. Note that $D$ lies within the quadrant $v_{1,2} \geq 1/2$.}
	\label{f:boundary}
	\end{figure}

The quantity $A$ defined in Eq. (\ref{e:A-B}) is a symmetric function of $v_1, v_2$ and $v_3$ which in turn are functions of $\eta$ and $\xi_2$ (\ref{e:conformal-prefactor-h}) for $0 \leq \eta \leq \pi/2$ and $0 \leq \xi_2 \leq \pi$. Since $\sum_i 1/v_i = 3$, we may regard $A$ as a function of any pair, say $v_1$ and $v_2$. The allowed values of $\eta$ and $\xi_2$ define a domain $\bar D = D \amalg \pdr D$ in the $v_1$-$v_2$ plane. To show that $A \geq 17/27$, we seek its global minimum, which must lie either at a local extremum in the interior $D$ or on the boundary $\pdr D$. $\pdr D$ is defined by the curves $\xi_2 = 0$ and $\xi_2 = \pi/2$ which meet at $\eta = 0$ and $\eta = \pi/2$ and include the points $(v_1 = \infty, v_2 = 2/3)$ and $(v_1 = 2/3, v_2 = \infty)$ (see Fig. \ref{f:boundary}). This is because, for any fixed $\eta$, $v_1$ and $v_2$ (\ref{e:conformal-prefactor-h}) are monotonic functions of $\xi_2$ for $0 \leq \xi_2 \leq \pi/2$ and symmetric under reflection about $\xi_2 = \pi/2$.
Along $\pdr D$, $A = (5 \cos 6 \eta + 22)/27$ is independent of $\xi_2$ and minimal at the Euler configurations $\eta = \pi/6$ and $\pi/2$ with the common minimum value $17/27$, which turns out to be the global minimum of $A$. This is because its only local extremum in $D$ is at the Lagrange configuration $v_1 = v_2 = v_3 = 1$ where $A = 2/3$. To see this, we note that local extrema of $A$ in $D$ must lie at the intersections of $\pdr A/\pdr v_1 = 0$ and $\pdr A/\pdr v_2 = 0$. Now $\pdr A/ \pdr v_1 = (v_1 - v_3)F(v_1, v_2)/v_1^2 u_2^3$ where
	\beq
	F(v_1,v_2)=  u_2 \left\{ v_1 + v_3 + 2 \left( v_1^2 + v_1 v_3 + v_3^2 \right) \right\} - 2 (v_1 + v_3) ( u_2 + u_4).
	\eeq
For $\pdr A/\pdr v_1$ to vanish, either $v_1 = v_3$ or $F(v_1,v_2) = 0$ or one of the $v_i = \infty$. The collision points $v_i = \infty$ do not lie in $D$. The conditions for $\pdr A / \pdr v_2$ to vanish are obtained via the exchange $v_1 \leftrightarrow v_2$. The intersection of the conditions $v_1 = v_3$ and $v_2 = v_3$ lies at the Lagrange configurations $v_i = 1$ where $A = 2/3$. It turns out that the only intersection of $v_1 = v_3$ with $F(v_2,v_1) = 0$ or of $v_2 = v_3$ with $F(v_1, v_2) = 0$ lying in $D$ occurs at the above Lagrange configuration. For instance, when $v_1 = v_3 = v$, $F(v_2,v_1) = -3 v^2 (4v-1)(v-1)/(3v - 2)^2$ vanishes when $v = 1$ or $v = 1/4$ (which violates $v \geq 1/2$). Finally, we account for extrema lying on the zero loci of both $F(v_1,v_2)$ and $F(v_2,v_1)$, which using $u_{-2} = 3$, must satisfy 
	\beq
	F(v_1,v_2) - F(v_2,v_1) = (v_1 - v_2) \left[12 v_1 v_2 v_3 - (v_1 + v_2 + v_3) \right] = 0.
	\eeq
So either $v_1 = v_2$ or $12 v_1 v_2 v_3 = u_2$. Now, we have shown above that the only extrema of $A$ on $v_1 = v_3$ in $D$ lie at the Lagrange configurations. Since $A$ is a symmetric function of the $v_i$, it follows that its only extrema on $v_1 = v_2$ also lies at the Lagrange configurations. On the other hand, $12 v_1 v_2 v_3 - (v_1 + v_2 + v_3) \geq 0$ for $v_i \geq 1/2$, with equality only at $v_i = 1/2$ which is not in $D$. Thus the only extremum of $A$ in $D$ is at the Lagrange configurations (where $A = 2/3$) and hence its global minimum occurs on $\pdr D$ at the Euler configurations  (where $A = 17/27$).

\small


\end{document}